\documentclass[12pt]{article}

\usepackage[a4paper,text={16.8cm,22.4cm}]{geometry}
\usepackage{amsmath,amsfonts,slashed,amssymb,tikz,bm,psfrag,graphicx,color,dsfont}
\usepackage{multicol}
\usepackage{float}

\RequirePackage[sort&compress,square,comma,numbers]{natbib}
\allowdisplaybreaks
\begin{document}

\begin{titlepage}

\begin{flushright}
\normalsize
 UWTHPH 2016-10 \\
June 9, 2016
\end{flushright}

\vspace{0.1cm}
\begin{center}
\Large\bf
Factorization and dispersion relations for radiative leptonic $B$ decay
\end{center}

\vspace{0.5cm}
\begin{center}
{\bf Yu-Ming Wang} \\
\vspace{0.7cm}
{\sl  ${}$\, Fakult\"{a}t f\"{u}r Physik, Universit\"{a}t Wien, Boltzmanngasse 5, 1090 Vienna, Austria \\
${}$\, School of Physics, Nankai University, 300071 Tianjin, China}
\end{center}

\vspace{0.2cm}
\begin{abstract}

Applying the dispersion approach we compute  perturbative QCD corrections to the power suppressed soft contribution of
$B \to \gamma \ell \nu$ at leading twist. QCD factorization for the $B \to \gamma^{\ast}$ form factors is
demonstrated explicitly for the  hard-collinear transverse polarized  photon at one loop,
with the aid of the method of regions. While the one-loop hard function  is identical to the
matching coefficient of the QCD weak current $\bar u \, \gamma_{\mu \perp} \, (1- \gamma_5) \, b$  in soft-collinear effective theory,
the jet function from integrating out the hard-collinear fluctuations differs from the corresponding one entering
the factorization formula of  $B \to \gamma \ell \nu$, due to the appearance of an additional hard-collinear momentum mode.
Furthermore, we evaluate the sub-leading power contribution to the $B \to \gamma$ form factors from the
three-particle $B$-meson distribution amplitudes (DAs) at tree level, with the dispersion approach.
The soft contribution to the $B \to \gamma$ form factors from the  three-particle $B$-meson DAs
is shown to be of the same power  compared with the  corresponding hard correction, in contrast to the two-particle counterparts.
Numerically the next-to-leading-order QCD correction to the soft two-particle contribution in $B \to \gamma$ form factors
will induce an approximately $\left (10 \sim 20 \right)$\% shift to the tree-level contribution
at $\lambda_B(\mu_0)=354 \, {\rm MeV}$. Albeit of  power suppression parametrically,
the soft two-particle correction can decrease the leading power
predictions for the $B \to \gamma$ form factors by an amount of $\left (10 \sim 30 \right)$\% with the same value of $\lambda_B(\mu_0)$.
Employing the phenomenological model of the three-particle $B$-meson DAs inspired
by a QCD sum rule analysis, the three-particle contribution to  the $B \to \gamma$ form factors
is predicted to be of ${\cal O} (1 \%)$, at leading order in $\alpha_s$, with the default theory inputs.
Finally, we explore  theory constraints on the inverse moment of the leading-twist $B$-meson DA $\lambda_B$
from  the recent Belle measurements of the partial branching fractions of $B \to \gamma \ell \nu$,
taking into account the newly computed contributions to the $B \to \gamma$ form factors
at subleading power.

\end{abstract}

\vfil

\end{titlepage}

\section{Introduction}
\label{sect:Intro}

The radiative leptonic $B \to \gamma \ell \nu$  decay serves as one of the benchmark channels
to understand the strong interaction dynamics of the $B$-meson system based upon the heavy quark expansion.
Factorization properties of the $B \to \gamma \ell \nu$  amplitude at large photon energy $E_{\gamma}$
have been explored extensively in both QCD \cite{Korchemsky:1999qb,DescotesGenon:2002mw}
and soft-collinear effective theory \cite{Lunghi:2002ju,Bosch:2003fc}
at leading power in $\Lambda/E_{\gamma}$. The particular feature of this channel lies in the strong sensitivity
of the branching faction ${\cal BR}(B \to \gamma \ell \nu)$ on the inverse moment $\lambda_B$ of the $B$-meson
light-cone distribution amplitude (DA) $\phi_B^{+}(\omega,\mu)$, which also enters the QCD factorization formulae for hadronic
$B$-meson decays. Improving the theory description of the radiative leptonic $B \to \gamma \ell \nu$  decay
by taking into account the subleading power effects is therefore  in demand to
achieve a better control over the inverse moment $\lambda_B$.

Subleading power corrections to $B \to \gamma \ell \nu$ in the heavy quark expansion  were
investigated in QCD factorization at tree level \cite{Beneke:2011nf} where a symmetry-conserving form factor
$\xi(E_{\gamma})$ was introduced to parameterize the non-local power correction.
It remains unclear whether $\xi(E_{\gamma})$ can be computed straightforwardly
in QCD factorization without encountering  rapidity divergences.
An alternative approach to evaluate the power suppressed contributions in $B \to \gamma \ell \nu$ was
proposed in \cite{Braun:2012kp} by employing the dispersion relations and quark-hadron duality,
where the ``soft" two-particle correction  to the $B \to \gamma$ form factors was computed at leading order
in the perturbative expansion. The main purpose of this paper  is to extend the calculation
performed in \cite{Braun:2012kp} by computing the subleading power contributions  to
the $B \to \gamma \ell \nu$ amplitude from the two-particle DA $\phi_B^{+}(\omega,\mu)$ at one loop
and from the three-particle DAs at tree level, for the sake of understanding the factorization properties of
the higher power terms in the heavy quark expansion.

The basic idea of the dispersion approach is to first construct the sum rules for
 the generalized form factors of $B \to \gamma^{\ast} \, \ell \nu$ involving
a spacelike hard-collinear photon with momentum $p$, and to perform the analytical continuation to $p^2=0$
to obtain the expressions for the on-shell $B \to \gamma$ form factors due to absence of the
massless vector resonances. The primary task of evaluating the two-particle contribution to the above-mentioned  sum rules
at next-to-leading order in $\alpha_s$ is to demonstrate QCD factorization  for the  $B \to \gamma^{\ast}$ form factors,
which can be achieved with either   the soft-collinear effective theory (SCET) technique
\cite{Bauer:2000yr,Bauer:2001yt,Beneke:2002ph} or  the diagrammatic approach  based upon the method of regions \cite{Beneke:1997zp}.
We will, following \cite{Wang:2015vgv,Wang:2015ndk}, adopt the latter approach to establish the factorization formula for
the leading-twist contribution to ${\cal A} (B \to \gamma^{\ast} \, \ell \nu)$
at one loop and employ the renormalization-group (RG) approach to resum large logarithms in the perturbative functions
at next-to-leading-logarithmic (NLL) accuracy.
It is evident that the hard function entering the factorization formula of ${\cal A} (B \to \gamma^{\ast} \, \ell \nu)$
with a (transversely polarized) hard-collinear photon can be extracted directly from the perturbative matching coefficient
of the  QCD weak current $\bar u \, \gamma_{\mu \perp} \, (1- \gamma_5) \, b$ in $\rm {SCET}$,
due to the absence of an additional hard-momentum mode; and  in the limit  $p^2=0$ the resulting hard-collinear function
must reproduce the jet function in the SCET factorization for the $B \to \gamma\, \ell \nu$ decay amplitude.
Applying the light-cone expansion for the massless quark propagator in the background gluon field,
we will demonstrate that QCD factorization for the three-particle contribution to ${\cal A} (B \to \gamma \, \ell \nu)$
is already violated  at tree level  due to the emergence of end-point divergences, and the dispersion approach
developed in  \cite{Braun:2012kp} provides a coherent framework to calculate the  subleading
power contributions from both the leading and higher Fock states of the $B$-meson.
Following the established power counting scheme, we further show that both
the ``hard" and ``soft" effects from the three-particle $B$-meson DAs
contribute to the sum rules at the same power in $\Lambda/m_b$, in contrast to the observation for the
leading twist contribution.

Yet another approach to address the subleading power contributions to the $B \to \gamma \, \ell \nu$
amplitude from the photon emission at large distance is to introduce the photon DAs describing
the strong interaction dynamics for the ``hadronic" component of a collinear real photon.
Employing the vacuum-to-photon correlation function with the $B$-meson replaced by a local pseudoscalar
current, the leading-twist contribution of such long-distance photon effect
has been computed from QCD light-cone sum rules (LCSR) at tree level \cite{Khodjamirian:1995uc,Eilam:1995zv}
and at one loop \cite{Ball:2003fq}. Interestingly,  the higher-twist correction to the hadronic
photon contribution calculated in the same framework was found to violate the symmetry relation for
two $B \to \gamma$ form factors due to the helicity conservation in the heavy quark limit \cite{Ball:2003fq}.
Computing the hadronic photon effect in ${\cal A} (B \to \gamma \, \ell \nu)$ from QCD factorization with the
photon DAs would be of great interest to develop a better understanding towards the pattern of the subleading
power contributions from different dynamical sources.
However,  it is quite conceivable that the convolution integral involving the $B$-meson and
photon DAs suffers from the end-point divergences, indicated  from a direct calculation
of the  similar effect on the $\pi \to \gamma$ form factor \cite{Agaev:2010aq}.

The presentation is structured as follows. In Section \ref{sect: dispersion approach}
we will discuss some general aspects of the  $B \to \gamma \ell \nu$ amplitude
and summarize the main idea of computing the (soft) end-point contributions to the $B \to \gamma$
form factors in the dispersion approach,
by working out the tree-level sum rules for the  power suppressed two-particle contribution.
We then demonstrate QCD factorization for the leading twist contribution
to the generalized  $B \to \gamma^{\ast} \ell \nu$ form factors
at ${\cal O}(\alpha_s)$ with the diagrammatical factorization approach
in Section \ref{sect:two-particle contribution}, where the sum rules for the
two-particle subleading power contribution to ${\cal A} (B \to \gamma \, \ell \nu)$
are also derived at NLL accuracy.
We further compute  the subleading power three-particle contribution to the $B \to \gamma$ form factors
from the dispersion approach at tree level in Section  \ref{sect:three-particle contribution},
which constitutes another new result of this paper.
Phenomenological implications of the newly computed contributions to the
$B \to \gamma \, \ell \nu$ amplitude are explored in Section \ref{sect:Numerical analysis},
including the uncertainty estimates for our predictions from different dynamical sources.
Section \ref{sect:Conc} is reserved for a summary of main observations and concluding discussions.

\section{The radiative leptonic $B \to \gamma \ell \nu$ decay in dispersion approach}
\label{sect: dispersion approach}

\subsection{General aspects of the $B \to \gamma \ell \nu$ amplitude}

We will follow closely the theory overview of $B \to \gamma \ell \nu$ presented in
\cite{Beneke:2011nf} and the corresponding decay amplitude can be written as
\begin{eqnarray}
{\cal A}(B^{-} \to \gamma \, \ell \, \nu )
=\frac{G_F \, V_{ub}} {\sqrt{2}} \, \left \langle \gamma(p) \, \ell(p_{\ell}) \, \nu(p_{\nu}) |
\left [ \bar{\ell} \, \gamma_{\mu} \, (1- \gamma_5) \, \nu  \right ] \,\,
\left [ \bar u \, \gamma^{\mu} \, (1- \gamma_5) \, b \right ] | B^{-}(p+q) \right \rangle  \,,
\label{def: decay amplitude}
\end{eqnarray}
where $p+q$ and $p$ denote the momenta of the $B$-meson and photon, and
the lepton-pair momentum is given by $q=p_{\ell}+p_{\nu}$
with $p_{\ell}$ and $p_{\nu}$ being the lepton and neutrino momenta, respectively.
We will work in the rest frame of the $B$-meson with the velocity vector  $v^{\mu}=(p^{\mu} + q^{\mu})/m_B$
and introduce two light-cone vectors $n_{\mu}$ and $\bar{n}_{\mu}$  by defining
\begin{eqnarray}
p_{\mu}=\frac{n \cdot p}{2}\, \bar{n}_{\mu} \equiv E_{\gamma} \, \bar{n}_{\mu}\,, \qquad
q_{\mu}=\frac{n \cdot q}{2}\, \bar{n}_{\mu} + \frac{\bar n \cdot q}{2}\, n_{\mu} \,,
\qquad v_{\mu} =\frac{ n_{\mu} + \bar{n}_{\mu}} {2} \,.
\end{eqnarray}

Computing the amplitude ${\cal A}(B^{-} \to \gamma \, \ell \, \nu )$ in (\ref{def: decay amplitude}) to
the first order in the electromagnetic interaction yields
\begin{eqnarray}
{\cal A}(B^{-} \to \gamma \, \ell \ \nu)
={G_F \, V_{ub} \over \sqrt{2}} \, \left ( i \, g_{em} \, \epsilon_{\nu}^{\ast}  \right )
\bigg \{ T^{\nu \mu}(p, q) \, \overline \ell \, \gamma_{\mu} \, (1-\gamma_5)  \nu
+ Q_{\ell} \,\, f_B \,\,\overline \ell \, \gamma^{\nu} \, (1-\gamma_5)  \nu  \bigg \} \,,
\label{original B to gamma l nu amplitude}
\end{eqnarray}
where the two terms in the bracket correspond to the photon emission from the partonic constitutes of the
$B$-meson and from the final-state lepton.
The hadronic tensor $T^{\nu \mu}(p, q)$ is defined by the
following non-local matrix element
\begin{eqnarray}
T_{\nu \mu}(p, q) &\equiv& \int d^4 x \, e^{i p \cdot x}  \,
\langle 0 | {\rm T} \{j_{\nu, \rm{em}}(x),
\left [\bar u \gamma_{\mu} (1-\gamma_5) b \right ] (0) \} |  B^{-}(p+q) \rangle \,,
\end{eqnarray}
where we adopt  the convention  for the QCD and QED covariant derivative
$i D_{\mu}= i \partial_{\mu} + g_{\rm em} \, Q_{f} A_{\mu, \rm {em}} + g_s \, T^a \, A_{\mu}^{a}$
with $Q_f=-1$ for the lepton fields,  and  the electromagnetic current is given by
$ j_{\nu, \rm{em}}=\sum_{q} Q_q \, \bar q \, \gamma_{\nu} \, q + Q_{\ell} \, \bar \ell \, \gamma_{\nu} \, \ell $.
It is straightforward to write down the general decomposition of this hadronic matrix element
\cite{Grinstein:2000pc,Khodjamirian:2001ga}
\begin{eqnarray}
T_{\nu \mu}(p, q) &=& v \cdot p \left [- i \, \epsilon_{\mu \nu \rho \sigma} \, n^{\rho} \, v^{\sigma} \, F_V(n \cdot p)
+ g_{\mu \nu} \, \hat{F}_A(n \cdot p)\right ] + v_{\nu} \, p_{\mu} \, F_1(n \cdot p)  \nonumber \\
&& + v_{\mu} \, p_{\nu} \, F_2(n \cdot p) + v \cdot p \,\, v_{\mu} \, v_{\nu} \, F_3(n \cdot p)
+\frac{p_{\mu} \, p_{\nu}}{v \cdot p } \, F_4(n \cdot p)   \,,
\end{eqnarray}
with the convention $\epsilon^{0123}=+1$.
It is evident that the form factors $F_2(n \cdot p)$ and $F_4(n \cdot p)$ will not contribute to the
amplitude ${\cal A}(B^{-} \to \gamma \, \ell \, \nu )$  in virtue of $\epsilon^{\ast} \cdot p=0$.
Employing the Ward identity $p_{\nu} \, T^{\nu \mu}(p, q) = -(Q_b-Q_u) \, f_B \, p_B^{\mu}$
due to the conservation of the vector current, we can further obtain
\begin{eqnarray}
\hat{F}_A(n \cdot p)= - F_1(n \cdot p) \,, \qquad F_3(n \cdot p)
= - \frac{(Q_b-Q_u) \, f_B \, m_B}{(v \cdot p)^2} \,.
\end{eqnarray}
Since the real photon is transversely polarized, the form factor $F_3(v \cdot p)$ will play no role
in the $B \to \gamma \, \ell \, \nu $ amplitude. Finally,  one can redefine the axial form factor
$\hat{F}_A(v \cdot p)$ \cite{Beneke:2011nf}
\begin{eqnarray}
T_{\nu \mu}(p, q) &\rightarrow& - i \, v \cdot p \, \epsilon_{\mu \nu \rho \sigma}
\, n^{\rho} \, v^{\sigma} \, F_V(n \cdot p)
+ \left [ g_{\mu \nu} \,v \cdot p - v_{\nu} \, p_{\mu} \right ] \,
\underbrace{\left [ \hat{F}_A(n \cdot p) +  \frac{Q_{\ell} \, f_B}{v \cdot p} \right ] } \nonumber \\
&&  -Q_{\ell} \, f_B \, g_{\mu \nu} \,, \hspace{7.0 cm}  \equiv F_A(n \cdot p)
\end{eqnarray}
where the last term cancels precisely the second term in the bracket of (\ref{original B to gamma l nu amplitude})
due to the photon radiation off the lepton.
The differential decay rate of $B \to \gamma \ell \nu$ in the $B$-meson rest frame can be readily computed as
\begin{eqnarray}
\frac{d \, \Gamma}{ d \, E_{\rm \gamma}} \left ( B \to \gamma \ell \nu \right )
=\frac{\alpha_{em}^2 \, G_F^2 \, |V_{ub}|^2}{6 \, \pi^2} \, m_B \, E_{\gamma}^3 \,
\left ( 1- \frac{2 \, E_{\gamma}}{m_B} \right ) \,
\left [ F_V^2(n \cdot p) + F_A^2(n \cdot p) \right ] \,.
\end{eqnarray}
Evaluating the partial branching fractions of  $B \to \gamma \ell \nu$
with an energetic photon is then traded to the QCD calculation of the two $B \to \gamma$ form factors.

\subsection{Dispersion relations for the $B \to \gamma$ form factors}

The aim of this subsection is to discuss the essential strategies for calculating
the $B \to \gamma$ form factors  from the dispersion approach
which is originally proposed in \cite{Khodjamirian:1997tk} for the computation of
the $\gamma^{\ast} \pi \to \gamma$ form factor with large momentum transfer
(see also \cite{Agaev:2010aq} for an updated  analysis including the higher-twist corrections).
Following \cite{Braun:2012kp} we start with construction of the correlation function
\begin{eqnarray}
\tilde{T}_{\nu \mu}(p, q) &\equiv& \int d^4 x \, e^{i p \cdot x}  \,
\langle 0 | {\rm T} \{ j_{\nu, \rm{em}}^{\perp}(x),
\left [\bar u \gamma_{\mu \, \perp} (1-\gamma_5) b \right ] (0) \} |  B^{-}(p+q) \rangle \big|_{p^2<0} \,, \nonumber \\
&=&  v \cdot p \, \left [ - i \, \epsilon_{\mu \nu \rho \sigma}
\, n^{\rho} \, v^{\sigma} \, F_V^{B \to \gamma^{\ast}}(n \cdot p, \bar n \cdot p)
+  g_{\mu \nu}^{\perp}  \, \hat{F}_A^{B \to \gamma^{\ast}} (n \cdot p, \bar n \cdot p) \right ]  \,,
\label{def: correlation function}
\end{eqnarray}
describing  the $B \to \gamma^{\ast} \ell \nu$ transition with a (transversely  polarized) hard-collinear photon.
For definiteness, we work with the following power counting scheme
\begin{eqnarray}
n \cdot p  \sim {\cal O}(m_b) \,, \qquad  |\bar n \cdot p| \sim {\cal O}(\Lambda)  \,.
\end{eqnarray}

\begin{figure}
\begin{center}
\includegraphics[width=0.60 \textwidth]{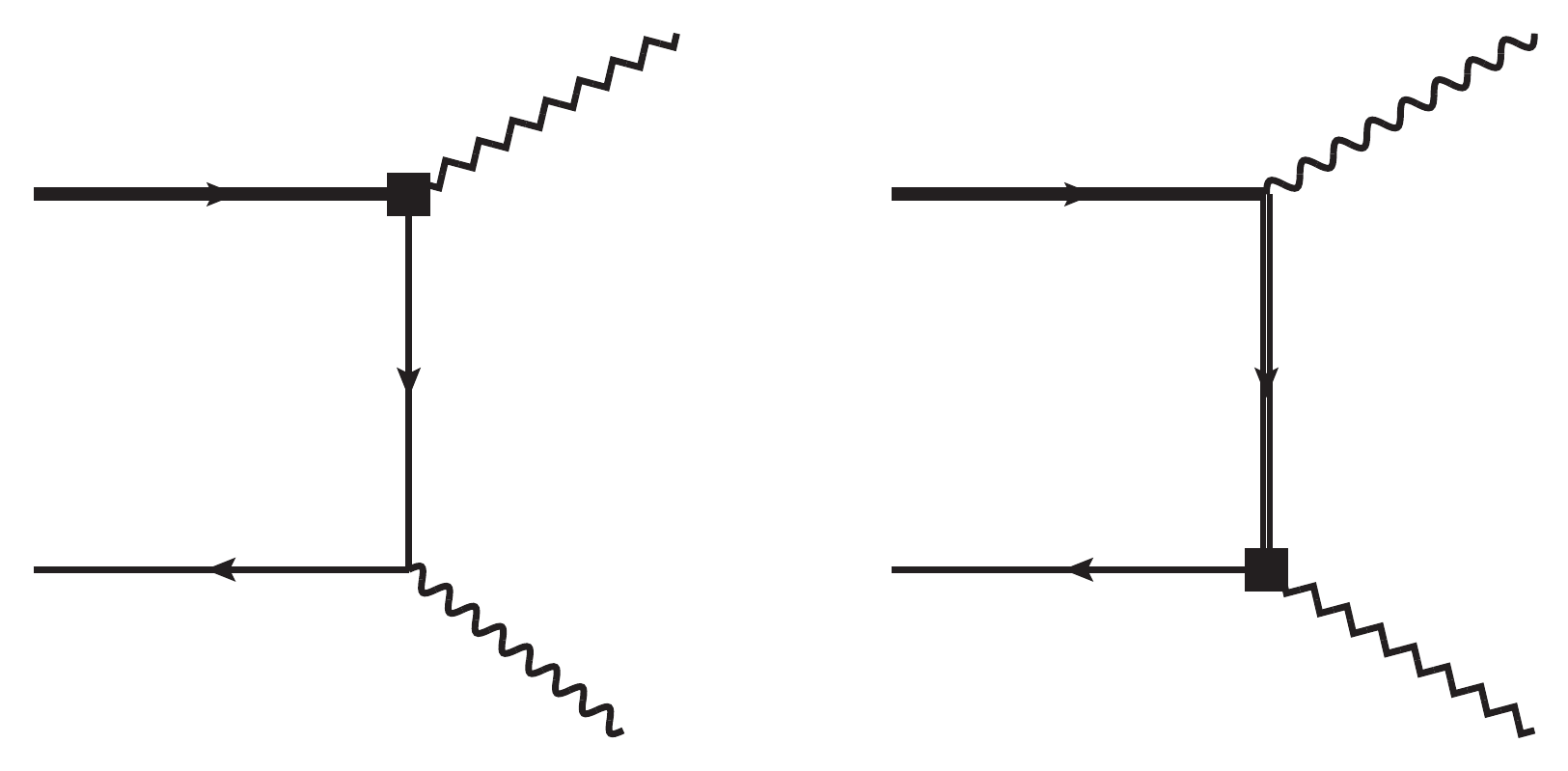}
\end{center}
\caption{Diagrammatical representation of the correlation function (\ref{def: correlation function})
at tree level. The square boxes refer to  insertions of the weak vertex ``$\bar u \gamma_{\mu \, \perp} (1-\gamma_5) b$"
and the wavelines indicate  photon radiations off the partons inside the $B$-meson. }
\label{tree diagram of the correlator}
\end{figure}

At tree level we need to evaluate the two diagrams displayed in figure \ref{tree diagram of the correlator}
with (light-cone) operator product expansion (OPE). It is apparent that
photon emission off the heavy $b$-quark will only induce the subleading power contribution
based upon the power counting analysis. The resulting local effect independent of the soft momentum $\bar n \cdot p$
is identical to the corresponding result presented in \cite{Beneke:2011nf}, however, the non-local subleading
power correction at tree level differs from the symmetry-conserving form factor $\xi(E_{\gamma})$
discussed in  the context of the $B \to \gamma \ell \nu$ transition.
In this paper we will take the Born result of the local subleading power contribution
to $\tilde{T}_{\nu \mu}(p, q)$ from \cite{Beneke:2011nf} directly
\begin{eqnarray}
F_{V, \, NLP}^{\rm {LC}}(n \cdot p)
= -\hat{F}_{A, \, NLP}^{\rm {LC}}(n \cdot p)
= \frac{Q_u \, f_B \, m_B}{(n \cdot p)^2}
+ \frac{Q_b \, f_B \, m_B}{n \cdot p \, m_b}  \,,
\label{subleading power local contribution}
\end{eqnarray}
and leave out the non-local power correction which could be expressed
in terms of the higher-twist $B$-meson DAs.

Computing the leading power contribution from photon radiation off the up anti-quark
at tree level yields
\begin{eqnarray}
&& F_{V, \,2P}^{B \to \gamma^{\ast}}(n \cdot p, \bar n \cdot p)
=\hat{F}_{A, \,2P}^{B \to \gamma^{\ast}}(n \cdot p, \bar n \cdot p) \nonumber \\
&& = \frac{Q_u \, \tilde{f}_B(\mu) \, m_B}{n \cdot p} \,
\, \int_0^{\infty} \, d \omega \, \frac{\phi_B^{+}(\omega, \mu)}{\omega - \bar n \cdot p - i 0} \,
+{\cal O}(\alpha_s, \Lambda/m_b) \,,
\label{correlator: QCD at tree level}
\end{eqnarray}
where the $B$-meson DA $\phi_B^{+}(\omega, \mu)$ is defined as \cite{Grozin:1996pq,Beneke:2000wa,Beneke:2005gs}
\begin{eqnarray}
i \, \tilde{f}_B(\mu) \, m_B \, \phi_B^{+}(\omega, \mu)
= {1 \over 2 \, \pi} \, \int_0^{\infty} d t \, e^{i \omega \, t } \,
\langle 0 | (\bar q_s \, Y_s)(t \, \bar n) \, \!  \not  {\bar n} \,  \gamma_5 \,
(Y_s^{\dag} \, b_v)(0) |  \bar B(v) \rangle  \,,
\end{eqnarray}
with the soft Wilson link
\begin{eqnarray}
Y_s(t \, \bar n)= {\rm P} \, \left \{ {\rm  Exp} \left [   i \, g_s \,
\int_{- \infty}^{t} \, dx \,  \bar n  \cdot A_{s}(x \, \bar n) \right ]  \right \} \,.
\end{eqnarray}
At one loop, the HQET decay constat $\tilde{f}_B(\mu)$ of the $B$-meson can be expressed in terms of
the QCD decay constant $f_B$ as follows
\begin{eqnarray}
\tilde{f}_B(\mu)= \left \{  1 +  {\alpha_s(\mu) \, C_F \over 4 \, \pi} \,
\left [3\, \ln{m_b \over \mu} -2  \right ] \right \}^{-1} \, f_B \,.
\label{matching condition for the fB}
\end{eqnarray}

Taking into account the fact that  $F_V^{B \to \gamma^{\ast}}$ and $\hat{F}_A^{B \to \gamma^{\ast}}$
are analytical functions in the variable $p^2$ (or $\bar n \cdot p$ equivalently), we can derive the
following hadronic dispersion relations
\begin{eqnarray}
F_V^{B \to \gamma^{\ast}}(n \cdot p, \bar n \cdot p)
&=&{2 \over 3} \, \frac{f_{\rho} \, m_{\rho}}{m_{\rho}^2-p^2-i 0}
\, {2 \, m_B \over m_B +  m_{\rho}} \, V(q^2) \nonumber \\
&& + {1 \over \pi} \, \int_{\omega_s}^{\infty} \, d \omega^{\prime} \,\,
\frac{{\rm Im}_{\omega^{\prime}} \, F_V^{B \to \gamma^{\ast}, \, {\rm had}}(n \cdot p, \omega^{\prime})}
{\omega^{\prime}- \bar n \cdot p - i 0}  \,, \,\,
\label{dispersion relation: FV}\\
\hat{F}_A^{B \to \gamma^{\ast}}(n \cdot p, \bar n \cdot p)
&=& {2 \over 3} \, \frac{f_{\rho} \, m_{\rho}}{m_{\rho}^2-p^2-i 0}
\, { 2 \left( m_B +  m_{\rho} \right ) \over n \cdot p } \, A_1(q^2) \nonumber \\
&& + {1 \over \pi} \, \int_{\omega_s}^{\infty} \, d \omega^{\prime} \,\,
\frac{{\rm Im}_{\omega^{\prime}} \, \hat{F}_A^{B \to \gamma^{\ast}, \, {\rm had}}(n \cdot p, \omega^{\prime})}
{\omega^{\prime}- \bar n \cdot p - i 0}  \,, \,\,
\label{dispersion relation: FAhat}
\end{eqnarray}
where the ground-state contributions from $\rho$ and $\omega$ are combined into one resonance term
with the narrow-width approximation and with the assumption $m_{\rho} \simeq m_{\omega}$.
The relevant $B \to \rho$ form factors are defined as
\begin{eqnarray}
\sqrt{2} \, \langle \rho^{0}(p) |\bar u \, \gamma_{\mu} \, (1-\gamma_5) \, b| B^{-}(p+q) \rangle
= - \epsilon_{\mu \nu \rho \sigma} \, \epsilon_V^{\ast \nu} \, q^{\rho} \, p^{\sigma} \,
\frac{2\, V(q^2)}{m_B+m_{\rho}} - i \, \epsilon_{V \mu}^{\ast} \, (m_B+m_{\rho}) \,  A_1(q^2) \nonumber \\
+ \, i \, (2 \, p+ q)_{\mu} \, \left( \epsilon_V^{\ast} \cdot q \right ) \,
\frac{A_2(q^2)}{m_B+m_{\rho}} - \, i \, q_{\mu} \left( \epsilon_V^{\ast} \cdot q \right ) \,
{2 \, m_{\rho} \over q^2} \, \left [ A_3(q^2)  - A_0(q^2) \right ] \,,  \hspace{0.4 cm}
\end{eqnarray}
where $\epsilon_V$ is the polarization vector of the $\rho$ meson.
Applying the parton-hadron duality approximation and performing the Borel transformation with respect to the
variable $\bar n \cdot p$ yields the sum rules for the  $B \to \rho$ form factors $V(q^2)$ and $A_1(q^2)$
\begin{eqnarray}
{2 \over 3} \, \frac{f_{\rho} \, m_{\rho}}{n \cdot p} \,
{\rm Exp} \left [-{m_{\rho}^2 \over n \cdot p \, \omega_M} \right ] \,
{2 \, m_B \over m_B +  m_{\rho}} \, V(q^2)
&=&  {1 \over \pi} \, \int_0^{\omega_s} \,\, d \omega^{\prime} \,\,  e^{-\omega^{\prime}/\omega_M} \,
\, {\rm Im}_{\omega^{\prime}} \, F_V^{B \to \gamma^{\ast}}(n \cdot p, \omega^{\prime})  \,, \hspace{0.8 cm}
\label{sum rules of the form factor V} \\
{2 \over 3} \, \frac{f_{\rho} \, m_{\rho}}{n \cdot p} \,
{\rm Exp} \left [-{m_{\rho}^2 \over n \cdot p \, \omega_M} \right ] \,
 { 2 \left( m_B +  m_{\rho} \right ) \over n \cdot p } \, A_1(q^2)
&=& {1 \over \pi} \, \int_0^{\omega_s} \,\, d \omega^{\prime} \,\,  e^{-\omega^{\prime}/\omega_M} \,
\,{\rm Im}_{\omega^{\prime}} \, \hat{F}_A^{B \to \gamma^{\ast}}(n \cdot p, \omega^{\prime})  \,, \hspace{0.8 cm}
\label{sum rules of the form factor A1}
\end{eqnarray}
where the QCD spectral functions at tree level can be readily extracted from (\ref{correlator: QCD at tree level})
\begin{eqnarray}
&& {1 \over \pi} \, {\rm Im}_{\omega^{\prime}} \, F_V^{B \to \gamma^{\ast}}(n \cdot p, \omega^{\prime}) =
{1 \over \pi} \, {\rm Im}_{\omega^{\prime}} \, \hat{F}_A^{B \to \gamma^{\ast}}(n \cdot p, \omega^{\prime}) \nonumber \\
&& = \frac{Q_u \, \tilde{f}_B(\mu) \, m_B}{n \cdot p} \, \phi_B^{+}(\omega^{\prime}, \mu) +{\cal O}(\alpha_s, \Lambda/m_b) \,.
\end{eqnarray}
Substituting the resulting LCSR (\ref{sum rules of the form factor V}) and (\ref{sum rules of the form factor A1})
into the dispersion relations (\ref{dispersion relation: FV}) and (\ref{dispersion relation: FAhat})
and setting $\bar n \cdot p \to 0$, we obtain the final expressions for the on-shell $B \to \gamma$  form factors
\begin{eqnarray}
F_V(n \cdot p) &=& {1 \over \pi} \, \int_0^{\omega_s} \,\, d \omega^{\prime} \,\,  \frac{n \cdot p}{m_{\rho}^2} \,
{\rm Exp} \left [{m_{\rho}^2 - \omega^{\prime} \, n \cdot p \over n \cdot p \, \omega_M} \right ]
\, \left [{\rm Im}_{\omega^{\prime}} \, F_V^{B \to \gamma^{\ast}}(n \cdot p, \omega^{\prime}) \right ]  \,  \nonumber \\
&& +  {1 \over \pi} \, \int_{\omega_s}^{\infty} \,\, d \omega^{\prime} \,\,  \frac{1}{\omega^{\prime}} \,
\, \left [{\rm Im}_{\omega^{\prime}} \, F_V^{B \to \gamma^{\ast}}(n \cdot p, \omega^{\prime}) \right ] \,,
\label{master formula of FV} \\
\hat{F}_A(n \cdot p) &=& {1 \over \pi} \, \int_0^{\omega_s} \,\, d \omega^{\prime} \,\,  \frac{n \cdot p}{m_{\rho}^2} \,
{\rm Exp} \left [{m_{\rho}^2 - \omega^{\prime} \, n \cdot p \over n \cdot p \, \omega_M} \right ]
\, \left [{\rm Im}_{\omega^{\prime}} \, \hat{F}_A^{B \to \gamma^{\ast}}(n \cdot p, \omega^{\prime}) \right ]  \,  \nonumber \\
&& +  {1 \over \pi} \, \int_{\omega_s}^{\infty} \,\, d \omega^{\prime} \,\,  \frac{1}{\omega^{\prime}} \,
\, \left [{\rm Im}_{\omega^{\prime}} \, \hat{F}_A^{B \to \gamma^{\ast}}(n \cdot p, \omega^{\prime}) \right ] \,,
\label{master formula of FAhat}
\end{eqnarray}
where two nonperturbative parameters $\omega_s$ and $m_{\rho}$ are introduced, as compared to the direct QCD calculation,
to avoid the evaluation of the ``hadronic" photon contribution.
To develop a better understanding of the master formulae  (\ref{master formula of FV})
and (\ref{master formula of FAhat}) for the form factors $F_V(n \cdot p)$
and $\hat{F}_A(n \cdot p)$, one can rewrite these expressions as follows
\begin{eqnarray}
F_V(n \cdot p) &=& {1 \over \pi} \, \int_{0}^{\infty} \,\,d \omega^{\prime} \,\,  \frac{1}{\omega^{\prime}} \,
\, \left [{\rm Im}_{\omega^{\prime}} \, F_V^{B \to \gamma^{\ast}}(n \cdot p, \omega^{\prime}) \right ]  \nonumber \\
&& + {1 \over \pi} \, \int_0^{\omega_s} \,\,d \omega^{\prime} \,\,  \left \{ \frac{n \cdot p}{m_{\rho}^2} \,
{\rm Exp} \left [{m_{\rho}^2 - \omega^{\prime} \, n \cdot p \over n \cdot p \, \omega_M} \right ]
- {1 \over \omega^{\prime}} \right \} \,
\, \left [{\rm Im}_{\omega^{\prime}} \, F_V^{B \to \gamma^{\ast}}(n \cdot p, \omega^{\prime}) \right ] \,,
\hspace{0.2 cm}  \label{modified master formula of FV} \\
\hat{F}_A(n \cdot p) &=& {1 \over \pi} \, \int_{0}^{\infty} \,\,d \omega^{\prime} \,\,  \frac{1}{\omega^{\prime}} \,
\, \left [{\rm Im}_{\omega^{\prime}} \, \hat{F}_A^{B \to \gamma^{\ast}}(n \cdot p, \omega^{\prime}) \right ]  \nonumber \\
&& + {1 \over \pi} \, \int_0^{\omega_s} \,\,d \omega^{\prime} \,\,  \left \{ \frac{n \cdot p}{m_{\rho}^2} \,
{\rm Exp} \left [{m_{\rho}^2 - \omega^{\prime} \, n \cdot p \over n \cdot p \, \omega_M} \right ]
- {1 \over \omega^{\prime}} \right \} \,
\, \left [{\rm Im}_{\omega^{\prime}} \, \hat{F}_A^{B \to \gamma^{\ast}}(n \cdot p, \omega^{\prime}) \right ] \,.
\label{modified master formula of FAhat}
\end{eqnarray}
It is evident that the first term on the right-hand side of (\ref{modified master formula of FV})
and (\ref{modified master formula of FAhat}) is precisely the  expression obtained from the
QCD factorization approach, provided that the convolution integrals of the
$B$-meson DAs with the perturbatively  calculable functions are free of  rapidity divergences.
In accordance with the power counting rule
\begin{eqnarray}
\omega_s \sim \omega_M  \sim {\cal O}(\Lambda^2/m_b) \,,
\label{power counting for SR parameters}
\end{eqnarray}
we observe that the second term on the right-hand side of (\ref{modified master formula of FV})
and (\ref{modified master formula of FAhat}) can be identified as the nonperturbative modification
of the spectral function in the end-point region. Exploring the canonical behaviour of the $B$-meson
DA  $\phi_B^{+}(\omega, \mu)$ and employing the tree-level expressions of the QCD spectral functions
(\ref{correlator: QCD at tree level}) one can readily verify that the end-point contributions to
the $B \to \gamma$ form factors are indeed suppressed by a factor of $\Lambda/m_b$ compared to the
effects computed from the direct QCD approach.

\section{Two-particle subleading power contribution at ${\cal O}(\alpha_s)$}
\label{sect:two-particle contribution}

\begin{figure}
\begin{center}
\includegraphics[width=0.90 \textwidth]{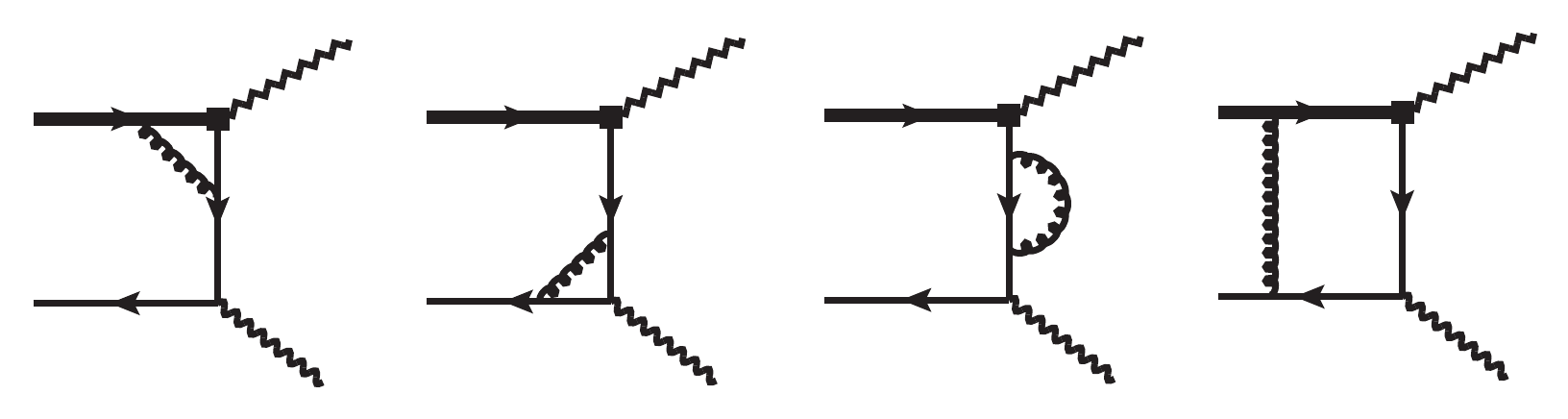} \\
(a) \hspace{3 cm} (b) \hspace{3 cm}  (c) \hspace{3 cm}   (d)
\end{center}
\caption{QCD corrections to the correlation function (\ref{def: correlation function})
at one loop. Same conventions as in figure \ref{tree diagram of the correlator}. }
\label{loop diagrams of the correlator}
\end{figure}

The objective of this section is to compute the one-loop corrections to
the perturbative hard and jet functions entering the factorization formulae of the generalized form factors
for the $B \to \gamma^{\ast} \ell \nu$ transition
\begin{eqnarray}
&& F_{V}^{B \to \gamma^{\ast}}(n \cdot p, \bar n \cdot p)
= \hat{F}_{A}^{B \to \gamma^{\ast}}(n \cdot p, \bar n \cdot p)  \nonumber  \\
&& = \frac{Q_u \, \tilde{f}_B(\mu) \, m_B}{n \cdot p} \, C_{\perp}(n \cdot p, \mu)  \,
\, \int_0^{\infty} \, d \omega \, \frac{\phi_B^{+}(\omega, \mu)}{\omega - \bar n \cdot p - i 0} \,
J_{\perp}(n \cdot p, \bar n \cdot p , \omega, \mu) + ... \,,
\label{master formulae for one-loop two particle factorization formula}
\end{eqnarray}
at leading power in $\Lambda/m_b$, where the ellipses represent the subleading power terms.
As mentioned in the Introduction,  we will extract the hard coefficient function ($C_{\perp}$)
and the jet function $J_{\perp}$ simultaneously by performing the perturbative matching
at the diagrammatic level  with the aid of the method of regions, and demonstrate the factorization-scale
independence of the form factors $F_{V}^{B \to \gamma^{\ast}}$ and $\hat{F}_{A}^{B \to \gamma^{\ast}}$
explicitly at one loop by exploiting the RG equations of both the short-distance functions
and the $B$-meson DA $\phi_B^{+}(\omega, \mu)$.
Now we will, following the presentation of \cite{Wang:2015vgv} closely, evaluate the one-loop QCD diagrams
for the correlation function (\ref{def: correlation function})
displayed figure \ref{loop diagrams of the correlator} in detail.

\subsection{Weak vertex diagram}

The one-loop correction to the weak vertex diagram displayed  in figure \ref{loop diagrams of the correlator}(a)
can be readily computed as
\begin{eqnarray}
\tilde{T}^{(1)}_{\nu \mu \,, weak}(p,q)
&=& \frac{Q_u \, g_s^2 \, C_F}{\bar n \cdot p -\omega} \, \int \frac{d^D l}{(2 \pi)^D} \,
\frac{1}{\left [(p-k+l)^2+ i 0 \right ] \, \left [(m_b \, v + l)^2 - m_b^2 + i 0 \right ]
\, \left [l^2 + i 0 \right ] } \nonumber \\
&& \times \left \{ n \cdot l  \left [ (D-2) \, \bar n \cdot l + 2\, m_b  \right ]   +  (D-4) \, l_{\perp}^2
+ 2\, n \cdot p \, (\bar n \cdot l + m_b)  \right \} \nonumber \\
&& \times  \, \bar u(k) \, \gamma_{\nu \perp} \, \frac{\not \! \bar n}{2} \, \gamma_{\mu \perp} \, \,(1 - \gamma_5) \, b(v)\,,
\label{full result of the weak vertex diagram}
\end{eqnarray}
where we adopt the following conventions
\begin{eqnarray}
l_{\perp}^2 \equiv g_{\perp}^{\mu \nu} \, l_{\mu} \, l_{\nu} \,, \qquad
g_{\perp}^{\mu \nu} \equiv g^{\mu \nu}-\frac{n^{\mu} \bar n^{\nu}}{2} -\frac{n^{\nu} \bar n^{\mu}}{2} \,.
\end{eqnarray}
Applying the power counting rule for the external momenta
\begin{eqnarray}
n \cdot p \sim {\cal O}(m_b) \,, \qquad  \bar n \cdot p \sim {\cal O}(\Lambda) \,, \qquad
k_{\mu} \sim {\cal O}(\Lambda)\,,
\label{power counting scheme}
\end{eqnarray}
it is straightforward to identify that the leading-power contributions of
$\tilde{T}^{(1)}_{\nu \mu \,, weak}$ arise from the hard, hard-collinear and soft regions
of the loop momentum.

Evaluating the leading power hard contribution from the weak vertex diagram  with the method of regions yields
\begin{eqnarray}
\tilde{T}^{(1),\,  h}_{\nu \mu \,, weak}(p,q)
&=& - i\, g_s^2 \, C_F \, \int \frac{d^D l}{(2 \pi)^D} \,
\frac{\tilde{T}^{(0)}_{\nu \mu}(p,q)}{\left [l^2 + n \cdot p \, \bar n \cdot l + i 0 \right ] \,
\left [l^2 + 2 \, m_b \, v \cdot l + i 0 \right ]
\, \left [l^2 + i 0 \right ] }   \nonumber \\
&& \times \left \{ n \cdot l  \left [ (D-2) \, \bar n \cdot l + 2\, m_b  \right ]   +  (D-4) \, l_{\perp}^2
+ 2\, n \cdot p \, (\bar n \cdot l + m_b)  \right \}  \,  \, \nonumber  \\
&\equiv& C_{\perp, weak}(n \cdot p) \,\,\,  \tilde{T}^{(0)}_{\nu \mu}(p,q) \,,
\end{eqnarray}
where $\tilde{T}^{(0)}_{\nu \mu}(p,q)$ is the leading order contribution to the correlation function
(\ref{def: correlation function})
\begin{eqnarray}
\tilde{T}^{(0)}_{\nu \mu}(p,q) = \frac{ i \, Q_u }{\bar n \cdot p -\omega}
\,\, \bar u(k) \, \gamma_{\nu \perp} \, \frac{\not \! \bar n}{2} \,
\gamma_{\mu \perp} \, \,(1 - \gamma_5) \, b(v) \,,
\end{eqnarray}
and the resulting hard function $C_{\perp, weak}(n \cdot p, \mu)$ is given by
\begin{eqnarray}
C_{\perp, weak}(n \cdot p, \mu) &=& -\frac{\alpha_s \, C_F}{4 \, \pi} \bigg [ {1 \over \epsilon^2} +
{1 \over \epsilon} \, \left ( 2 \, \ln {\mu \over  n \cdot p} + 1  \right )
+ 2 \, \ln^2 {\mu \over  n \cdot p} + 2\, \ln {\mu \over m_b}
 \nonumber \\
&& -2 \,  {\rm Li_2} \left (1- {1 \over r} \right ) - \ln^2 r + \frac{3 r-2}{1-r} \, \ln r
+\frac{\pi^2}{12} + 4  \bigg ]\,,
\end{eqnarray}
with $r=n \cdot p/m_b$. It is evident that $C_{\perp, weak}(n \cdot p)$ is precisely the same
as the hard contribution to the weak vertex diagram for the vacuum-to-$\Lambda_b$-baryon
correlation function  at one loop \cite{Wang:2015ndk}.

Proceeding in a similar manner, we can extract the hard-collinear correction from
figure \ref{loop diagrams of the correlator}(a) by expanding (\ref{full result of the weak vertex diagram})
in the hard-collinear region
\begin{eqnarray}
\tilde{T}^{(1),\,  hc}_{\nu \mu \,, weak}(p,q)
&=& - i\, g_s^2 \, C_F \, \int \frac{d^D l}{(2 \pi)^D} \,
\frac{2 \, m_b \, n \cdot (p+l) \,\, \tilde{T}^{(0)}_{\nu \mu}(p,q) }
{[ n \cdot (p+l) \,\bar n \cdot (p-k+l) + l_{\perp}^2  + i 0][ m_b \, n \cdot l+ i 0] [l^2+i0] }  \,\, \nonumber \\
& \equiv & J_{\perp, weak}(n \cdot p, \bar n \cdot p, \omega) \,\,\,  \tilde{T}^{(0)}_{\nu \mu}(p,q) \,,
\end{eqnarray}
where the perturbative jet function $J_{\perp, weak}(n \cdot p, \bar n \cdot p, \omega)$ at ${\cal O}(\alpha_s)$ reads
\begin{eqnarray}
J_{\perp, weak}(n \cdot p, \bar n \cdot p, \omega, \mu) &=&  \frac{\alpha_s \, C_F}{4 \, \pi}
\bigg [ {2 \over \epsilon^2} + {2 \over \epsilon} \,
\left (\ln {\mu^2 \over  n \cdot p \, (\omega - \bar n \cdot p)} + 1  \right )
 + \ln^2 {\mu^2 \over  n \cdot p \, (\omega - \bar n \cdot p)} \, \nonumber \\
&&  + \, 2 \, \ln {\mu^2 \over  n \cdot p \, (\omega - \bar n \cdot p)}  
-{\pi^2 \over 6} + 4 \bigg ] \,,
\end{eqnarray}
in agreement with \cite{Wang:2015vgv}. Setting $\bar n \cdot  p \to 0$, our result of $J_{\perp, weak}$
reproduces the hard-collinear contribution to the weak vertex diagram in the $B \to \gamma \ell \nu$
decay (see (69) in \cite{DescotesGenon:2002mw}).

Furthermore, expanding the full QCD amplitude of $\tilde{T}^{(1)}_{\nu \mu \,, weak}(p,q)$
in the soft region at leading power leads to the soft contribution
\begin{eqnarray}
\tilde{T}^{(1),\,  s}_{\nu \mu \,, weak}(p,q)
&=& - i\, g_s^2 \, C_F \, \int \frac{d^D l}{(2 \pi)^D} \,
\frac{1}{\left [ \bar n \cdot (p-k+l) + i 0 \right ] \left [ v \cdot l+ i 0 \right ] \left [l^2+i0 \right ] }
\,\, \tilde{T}^{(0)}_{\nu \mu}(p,q)   \,\,, \,\,
\end{eqnarray}
which cancels precisely the soft subtraction term defined by the convolution integral
of the two-particle $B$-meson DA $\phi_B^{+}(\omega, \mu)$ at ${\cal O}(\alpha_s)$ with the tree-level hard scattering kernel.
One then concludes that  soft dynamics of the weak vertex diagram in figure \ref{loop diagrams of the correlator}(a)
can indeed be parametrized by the $B$-meson DAs in the framework of perturbative QCD.

\subsection{Electromagnetic vertex diagram}

We proceed to compute the one-loop correction to the electromagnetic vertex diagram shown in figure
\ref{loop diagrams of the correlator}(b)
\begin{eqnarray}
\tilde{T}^{(1)}_{\nu \mu, {\rm em}}(p, q) = \frac{Q_u \, g_s^2 \, C_F}{n \cdot p \,\,  (\omega -\bar n \cdot p)}  \,
\int \frac{d^D l}{(2 \, \pi)^D}
 \frac{1}{[l^2 + i 0] [(p-l)^2+ i 0] [(l-k)^2 + i 0]} && \nonumber \\
\bar u(k) \,\, \gamma_{\rho} \,\, \! \not l \,\, \gamma_{\nu}^{\perp} \,\, (\!  \not p - \!  \not l )\,\,
\gamma^{\rho} \,\, (\!  \not p - \!  \not k ) \,\, \gamma_{\mu}^{\perp} \,\,
(1-\gamma_5)  \,\, b(v) \,.
\label{full result of the em vertex}
\end{eqnarray}
Employing the power counting rule (\ref{power counting scheme}) one can verify that only the hard-collinear
and soft regions in (\ref{full result of the em vertex}) can give rise to the leading power contributions.
Following the arguments of computing the pion vertex diagram  for
the vacuum-to-$B$-meson correlation function \cite{Wang:2015vgv}, it is more transparent to compute the loop integrals
in (\ref{full result of the em vertex}) directly instead of employing the method of regions, and then to expand the
resulting expression to the leading power in $\Lambda/m_b$.
Evaluating the loop integral with the expressions collected in Appendix A of \cite{Wang:2015vgv} yields
\begin{eqnarray}
\tilde{T}^{(1)}_{\nu \mu, {\rm em}}(p, q) &=&
{\alpha_s \, C_F \over 4 \, \pi} \,
\bigg  \{ {1 \over \epsilon} \, \left [ {2 \over \eta} \, \ln \left( 1+\eta \right ) - 1 \right ]
+ { \ln (1+ \eta)\over \eta} \,
\left [ 2 \, \ln { \mu^2 \over -p^2} - \ln (1+ \eta) + 3 \right ] \, \nonumber \\
&& -  \ln { \mu^2 \over n \cdot p \, (\omega - \bar n \cdot p)}  - 4  \bigg \}
\,\,\, \tilde{T}^{(0)}_{\nu \mu}(p, q) \nonumber \\
& \equiv & J_{\perp, {\rm em}} (n \cdot p, \bar n \cdot p, \omega, \mu) \, \, \tilde{T}^{(0)}_{\nu \mu}(p, q)  \,,
\end{eqnarray}
with $\eta= - \omega / \bar n \cdot p $.
It is straightforward to confirm that the obtained jet function $J_{\perp, {\rm em}}$
can reproduce the  hard-collinear correction to  the electromagnetic vertex diagram in $B \to \gamma \ell \nu$
(see (33) in \cite{DescotesGenon:2002mw} and (A.5) in \cite{Lunghi:2002ju}) in the limit $\bar n \cdot p \to 0$,
taking into account the fact that the soft contribution to $\tilde{T}^{(1)}_{\nu \mu, {\rm em}}(p, q)$  vanishes
in dimensional regularization. Following \cite{Wang:2015vgv} one can further verify that the soft contribution from the
electromagnetic vertex diagram cancels precisely the corresponding soft subtraction term, independent of the regularization
scheme.

\subsection{Wave function renormalization}

The contribution from the wave function renormalization of the immediate quark propagator in
figure \ref{loop diagrams of the correlator}(c) can be readily computed as
\begin{eqnarray}
\tilde{T}^{(1)}_{\nu \mu, {wfc}}(p, q) &=&
-{\alpha_s \, C_F \over 4 \, \pi} \,
\left [{1 \over \epsilon}
+ \ln {\mu^2 \over n \cdot p  \, \left ( \omega -\bar n \cdot p  \right ) }  + 1 \right ]
\, \, \tilde{T}^{(0)}_{\nu \mu}(p, q) \nonumber \\
& \equiv & J_{\perp, wfc} (n \cdot p, \bar n \cdot p, \omega, \mu) \, \, \tilde{T}^{(0)}_{\nu \mu}(p, q)  \,,
\end{eqnarray}
which is apparently free of soft and collinear divergences.
Evaluating the perturbative matching coefficients from the wave function renormalization
of the external quark fields yields
\begin{eqnarray}
\tilde{T}^{(1)}_{\nu \mu, {bwf}}(p, q) - \Phi_{b \bar u, bwf}^{(1)} \otimes \tilde{T}^{(0)}_{\nu \mu}(p, q)
&=&  - \frac{\alpha_s \, C_F}{8 \, \pi} \,
\bigg [{3 \over \epsilon} + 3 \, \ln {\mu^2 \over m_b^2} + 4 \bigg ] \, \tilde{T}^{(0)}_{\nu \mu}(p, q) \nonumber \\
& \equiv & C_{\perp, bwf}(n \cdot p, \mu) \, \tilde{T}^{(0)}_{\nu \mu}(p, q) \,,   \\
\tilde{T}^{(1)}_{\nu \mu, {uwf}}(p, q) - \Phi_{b \bar u, bwf}^{(1)} \otimes \tilde{T}^{(0)}_{\nu \mu}(p, q) &=&  0\,,
\end{eqnarray}
where $\Phi_{b \bar u}$ is the partonic DA of the $B$-meson defined in (12) of \cite{Wang:2015vgv}.

\subsection{Box diagram}

Now we turn to compute the one-loop contribution to $\tilde{T}^{(1)}_{\nu \mu}(p, q)$
from the box diagram shown in figure \ref{loop diagrams of the correlator}(d)
\begin{eqnarray}
\Pi_{\mu,  \, box}^{(1)}
&=& - Q_u \, g_s^2 \, C_F \,
\int \frac{d^D \, l}{(2 \pi)^D} \,   \frac{1}{[(m_b v+l)^2 -m_b^2+ i 0][(p-k+l)^2 + i 0] [(k-l)^2+i0][l^2+i0]}  \nonumber  \\
&& \bar u(k)  \, \gamma_{\rho}  \,  \,\,  (\! \not k - \! \not l) \,\,
\gamma_{\nu \perp}   \,\,  (\! \not p - \! \not k  + \! \not l) \, \gamma_{\mu \perp} \, (1-\gamma_5)\,\!
\, (m_b  \! \not v +  \! \not l+ m_b )\, \gamma^{\rho} \, b(v)  \,.
\label{diagram d: expression}
\end{eqnarray}
As discussed in \cite{DescotesGenon:2002mw},  this is the only one-loop diagram with no hard-collinear propagator outside
of the loop, therefore the $1 / (\omega - \bar n \cdot p)$ enhancement factor observed
in the tree-level result (\ref{correlator: QCD at tree level}) must come from singular regions
of phase space in the loop  integral. Based upon the power counting analysis, one can identify that the hard-collinear
contribution to the following four-point scalar integral
\begin{eqnarray}
I_{box} =\int \frac{d^D \, l}{(2 \pi)^D} \,
\frac{1}{[(m_b v+l)^2 -m_b^2+ i 0][(p-k+l)^2 + i 0] [(k-l)^2+i0][l^2+i0]}
\label{scalar integral: box diagram}
\end{eqnarray}
scales as $\lambda^{-1}$ with the expansion parameter $\lambda = \Lambda/m_b$.
The one-loop box diagram would then generate non-vanishing contribution to the
jet function $J_{\perp}(n \cdot p, \bar n \cdot p , \omega, \mu)$ entering the factorization formula
(\ref{def: decay amplitude}), provided that no additional suppression factor of $\lambda$
can be induced from the Dirac algebra in (\ref{diagram d: expression}).
Inspecting the Dirac structure in the numerator of  (\ref{diagram d: expression})
\begin{eqnarray}
(\! \not k - \! \not l) \,\, \gamma_{\nu \perp}   \,\,  (\! \not p - \! \not k  + \! \not l) \, \nonumber
\end{eqnarray}
shows that one cannot pick up the leading contributions of two hard-collinear  propagators
simultaneously in contrast to the case of the vacuum-to-$B$-meson correlation function
as considered in \cite{Wang:2015vgv}. Hence, one can conclude that no hard-collinear contribution
can arise from the box diagram at one loop displayed in figure \ref{loop diagrams of the correlator}(d),
confirming the observation  made in \cite{DescotesGenon:2002mw}.

Along the same vein, one can verify that  the soft contribution to
the scalar integral (\ref{scalar integral: box diagram}) scales as $\lambda^{-2}$ and the Dirac algebra
in (\ref{diagram d: expression}) will again give rise to a suppression factor of $\lambda$ in the soft region.
It is then evident that the leading-power contribution to the box diagram comes only from the soft region
at one loop, and following \cite{Wang:2015vgv}, such soft contribution will be cancelled precisely by
the corresponding infrared substraction term  from the standard perturbative matching procedure.

\subsection{Factorization of the two-particle contribution at ${\cal O}(\alpha_s)$}

Collecting everything together, we can readily derive the renormalized hard and jet functions entering the
factorization formula (\ref{master formulae for one-loop two particle factorization formula})
for the generalized $B \to \gamma^{\ast}$ form factors at one loop
\begin{eqnarray}
 C_{\perp} &=&  1 + C_{\perp, weak} + C_{\perp, bwf}  \nonumber \\
&=& 1- \frac{\alpha_s \, C_F}{4 \, \pi}
\bigg [ 2 \, \ln^2 {\mu \over n \cdot p} + 5 \, \ln {\mu \over m_b}
-2 \, {\rm Li}_2 \left ( 1-{1 \over r} \right )  - \ln^2  r  \nonumber \\
&&  + \,  {3 r -2 \over 1 -r}  \, \ln r + {\pi^2 \over 12} + 6  \bigg ] \,,
\label{one loop hard function}  \\
J_{\perp} &=& 1 + J_{\perp, weak}
+ J_{\perp, {\rm em}} + J_{\perp, wfc}  \nonumber \\
&=& 1 + {\alpha_s \, C_F \over 4 \, \pi} \,
\bigg \{ \ln^2 { \mu^2 \over n \cdot p \,  (\omega - \bar n \cdot p)}  - {\pi^2 \over 6} - 1  \nonumber \\
&& - {\bar n \cdot p \over \omega} \, \ln {\bar n \cdot p - \omega \over \bar n \cdot p } \,
\left [ \ln { \mu^2 \over -p^2} + \ln { \mu^2 \over n \cdot p \,  (\omega - \bar n \cdot p)} + 3 \right ]  \bigg \}  \,.
\label{one loop jet function}
\end{eqnarray}
It is straightforward to verify that the hard function $ C_{\perp}$ coincides with the perturbative
matching coefficient of the QCD weak current $\bar u \, \gamma_{\mu \perp} (1-\gamma_5) b\, $
in  SCET \cite{Bauer:2000ew,Beneke:2004rc}
\begin{eqnarray}
\bar u \, \gamma_{\mu \perp} \, (1-\gamma_5) \,  b \to C_3(\mu) \,
\bar \xi_{\bar n} \,W_{hc} \, \gamma_{\mu \perp} \, (1-\gamma_5) \, \, Y_{s}^{\dag} \, b_v + ... \,,
\end{eqnarray}
where $W_{hc}$  refers to the hard-collinear Wilson line and
the ellipses represent the subleading power contributions.

We are now in a position to demonstrate the factorization-scale independence of the factorization
formulae for $F_{V}^{B \to \gamma^{\ast}}$ and $\hat{F}_{V}^{B \to \gamma^{\ast}}$ explicitly at one loop.
With the expressions for the hard and jet functions in (\ref{one loop hard function})
and (\ref{one loop jet function}), we obtain
\begin{eqnarray}
&& \frac{d}{d \ln \mu} F_{V}^{B \to \gamma^{\ast}}
= \frac{d}{d \ln \mu} \hat{F}_{V}^{B \to \gamma^{\ast}} \nonumber \\
&& = \frac{Q_u \, \tilde{f}_B(\mu) \, m_B}{n \cdot p} \,
\bigg \{ \int_0^{\infty} \, d \omega \, \frac{\phi_B^{+}(\omega, \mu)}{\omega - \bar n \cdot p - i 0}  \,
\frac{\alpha_s \, C_F}{4 \, \pi} \,\,
\bigg[ - \left ( 4 \, \ln {\mu \over  n \cdot p }  + 5 \right )  \nonumber \\
&& \hspace{0.5 cm} + \,\, 4 \, \left ( \ln {\mu^2 \over n \cdot p \, (\omega- \bar n \cdot p) }
- {\bar n \cdot p \over \omega} \, \ln {\bar n \cdot p - \omega \over \bar n \cdot p} \right )
+ 3  \bigg ] \nonumber \\
&& \hspace{0.5 cm}  + \int_0^{\infty} \, d \omega \, \frac{1}{\omega - \bar n \cdot p - i 0} \,\,\,
\frac{d}{d \ln \mu} \, \phi_B^{+}(\omega, \mu)  \bigg \}\, \,,
\label{scale independence: original form}
\end{eqnarray}
where the three terms in the square bracket appeared on the right-handed side
of (\ref{scale independence: original form}) correspond to the contributions from
the scale evolutions of the hard and jet functions as well as the HQET decay constant
of the $B$-meson, respectively. Employing the one-loop evolution equation
of $\phi_B^{+}(\omega, \mu)$ \cite{Lange:2003ff,Braun:2003wx}
\begin{eqnarray}
\frac{d \phi_B^{+}(\omega, \mu)}{d \ln \mu} \,
= - \left [ \Gamma_{\rm {cusp}}(\alpha_s) \, \ln{\mu \over \omega}
+ \gamma_{+}(\alpha_s)  \right ] \, \phi_B^{+}(\omega, \mu)
- \omega \, \int_0^{\infty} \, d \omega^{\prime} \,
\Gamma_{+}(\omega,\omega^{\prime},\mu) \, \phi_B^{+}(\omega^{\prime}, \mu)
\label{Lange-Neubert equation}
\end{eqnarray}
with the anomalous dimensions
\begin{eqnarray}
\Gamma_{\rm {cusp}}(\alpha_s) &=& \sum_{n=0} \, \left ( \frac{\alpha_s}{4 \, \pi} \right )^{n+1} \,
\Gamma_{\rm {cusp}}^{(n)} \,,  \qquad   \Gamma_{\rm {cusp}}^{(0)} = 4 \, C_F \,, \nonumber \\
\gamma_{+}(\alpha_s) &=& \sum_{n=0} \, \left ( \frac{\alpha_s}{4 \, \pi} \right )^{n+1} \,
\gamma_{+}^{(n)} \,, \qquad   \gamma_{+}^{(0)} = - 2 \, C_F \,, \nonumber \\
\Gamma_{+}(\omega,\omega^{\prime},\mu) &=& -  \frac{\alpha_s}{4 \, \pi} \, \Gamma_{\rm cusp}^{(0)} \,
\left [ \frac{\theta(\omega^{\prime}-\omega)}{\omega^{\prime} \, (\omega^{\prime}-\omega)}
+ \frac{\theta(\omega-\omega^{\prime})}{\omega \, (\omega-\omega^{\prime})}  \right ]_{\oplus}
+ {\cal O}(\alpha_s^2) \,,
\end{eqnarray}
the last term in the bracket of (\ref{scale independence: original form}) can be further computed as
\begin{eqnarray}
&& \int_0^{\infty} \, d \omega \, \frac{1}{\omega - \bar n \cdot p - i 0} \,\,\,
\frac{d}{d \ln \mu} \, \phi_B^{+}(\omega, \mu) \nonumber \\
&& = \frac{\alpha_s \, C_F}{4 \, \pi} \, \int_0^{\infty} \, d \omega \,
\frac{\phi_B^{+}(\omega, \mu)}{\omega - \bar n \cdot p - i 0}  \,\,
\left [  4 \, \left ( - \ln {\mu  \over  \omega- \bar n \cdot p }
+ {\bar n \cdot p \over \omega} \, \ln {\bar n \cdot p - \omega \over \bar n \cdot p} \right )
+ 2 \right ]  \,.
\label{B-meson evolution contribution}
\end{eqnarray}
Here, the ``$\oplus$"-function is defined in a standard way
\begin{eqnarray}
\int_0^{\infty} \, d \omega \, \left [ f(\omega,\omega^{\prime}) \right ]_{\oplus} \, g(\omega)
= \int_0^{\infty} \, d \omega \, f(\omega,\omega^{\prime})  \,
\left [ g(\omega) - g(\omega^{\prime}) \right ] \,.
\end{eqnarray}
Substituting (\ref{B-meson evolution contribution}) into (\ref{scale independence: original form}) leads us to
conclude that the factorization-scale dependence indeed cancels out in the factorization formulae
for the $B \to \gamma^{\ast}$ form factors at one loop, i.e.,
\begin{eqnarray}
&& \frac{d}{d \ln \mu} F_{V}^{B \to \gamma^{\ast}}
= \frac{d}{d \ln \mu} \hat{F}_{A}^{B \to \gamma^{\ast}} = {\cal O}(\alpha_s^2) \,.
\end{eqnarray}

Now we turn to sum  the parametrically large logarithms in perturbative matching coefficients
to all orders at NLL by employing the standard RG approach in momentum space.
Since the hard-collinear scale $\mu_{hc} \simeq \sqrt{\Lambda \, m_b}$ is comparable to the soft scale
$\mu_0$ entering the initial condition of the $B$-meson DA $\phi_B^{+}(\omega, \mu_0)$ for the actual
value of the $b$-quark mass, we will not sum logarithms of $\mu_{hc} / \mu_0$ from the RG evolution
of $\phi_B^{+}(\omega, \mu)$  when the factorization scale $\mu$ is taken as a hard-collinear scale
as argued in \cite{Beneke:2011nf}. Solving the evolution equations for the hard function $C_{\perp}$
and the HQET decay constant $\tilde{f}_B$
\begin{eqnarray}
\frac{d C_{\perp}(n \cdot p, \mu)}{d \ln \mu} \,
&=& \left [ - \Gamma_{\rm {cusp}}(\alpha_s) \, \ln{\mu \over n \cdot p}
+ \gamma(\alpha_s)  \right ] \, C_{\perp}(n \cdot p, \mu) \,,  \nonumber \\
\frac{d \tilde{f}_B(\mu)}{d \ln \mu} \,
&=&  \tilde{\gamma}(\alpha_s) \, \tilde{f}_B(\mu) \,,
\end{eqnarray}
with the cusp anomalous dimension $\Gamma_{\rm {cusp}}(\alpha_s)$ expanded up to three loops and the remaining
anomalous dimensions $\gamma(\alpha_s)$ and $\tilde{\gamma}(\alpha_s)$ expanded up to the two-loop order, we then
obtain the NLL resummation improved expressions for $C_{\perp}$ and $\tilde{f}_B$
\begin{eqnarray}
C_{\perp}(n \cdot p, \mu) &=& U_1(n \cdot p, \mu_{h1}, \mu) \, C_{\perp}(n \cdot p, \mu_{h1}) \,, \nonumber \\
\tilde{f}_B(\mu) &=&  U_2(n \cdot p, \mu_{h2}, \mu) \,\tilde{f}_B(\mu_{h2}) \,.
\end{eqnarray}
The manifest expression of $U_1(n \cdot p, \mu_{h1}, \mu)$ can be deduced from
$U_1(E_{\gamma}, \mu_{h1}, \mu)$ in \cite{Beneke:2011nf} by replacing $E_{\gamma} \to n \cdot p/2$
and $ U_2(n \cdot p, \mu_{h2}, \mu) $ can be read from $U_1(E_{\gamma}, \mu_{h1}, \mu)$ by
setting the cusp anomalous dimension to zero and by replacing $\gamma^{(n)} \to \tilde{\gamma}^{(n)}$.

Finally we present the factorization formulae for the form factors $F_{V}^{B \to \gamma^{\ast}}$
and $\hat{F}_{A}^{B \to \gamma^{\ast}}$ with RG improvement at NLL accuracy
\begin{eqnarray}
&& F_{V}^{B \to \gamma^{\ast}}(n \cdot p, \bar n \cdot p)
= \hat{F}_{A}^{B \to \gamma^{\ast}}(n \cdot p, \bar n \cdot p)  \nonumber  \\
&& = \frac{Q_u \, m_B}{n \cdot p} \, \left [  U_2(n \cdot p, \mu_{h2}, \mu) \,\tilde{f}_B(\mu_{h2}) \right ] \,
\left [ U_1(n \cdot p, \mu_{h1}, \mu) \, C_{\perp}(n \cdot p, \mu_{h1})   \right ] \nonumber \\
&& \hspace{0.5 cm} \times \,\int_0^{\infty} \, d \omega \,
 \frac{\phi_B^{+}(\omega, \mu)}{\omega - \bar n \cdot p - i 0} \,
J_{\perp}(n \cdot p, \bar n \cdot p , \omega, \mu) + ... \,,
\label{resummation improved factorization formula}
\end{eqnarray}
where the factorization scale needs to be chosen as a hard-collinear scale of order $\sqrt{\Lambda \, m_b}$,
and $\mu_{h1}$ and $\mu_{h2}$ are the hard scales of order $m_b$.

\subsection{Dispersion relation for the two-particle contribution at ${\cal O}(\alpha_s)$}

It is now a straightforward task to derive the NLL resummation improved dispersion relations for the on-shell
$B \to \gamma$ form factors. Employing the spectral representations of various convolution integrals displayed
in Appendix \ref{app:spectral resp} yields
\begin{eqnarray}
&& F_{V, 2P}(n \cdot p)
= \hat{F}_{A,2P}(n \cdot p)  \nonumber  \\
&& = \frac{Q_u \, m_B}{n \cdot p} \, \left [  U_2(n \cdot p, \mu_{h2}, \mu) \,\tilde{f}_B(\mu_{h2}) \right ] \,
\left [ U_1(n \cdot p, \mu_{h1}, \mu) \, C_{\perp}(n \cdot p, \mu_{h1})   \right ] \nonumber \\
&& \hspace{0.5 cm} \times \, \bigg \{   \,\int_0^{\infty} \, d \omega \,
 \frac{\phi_B^{+}(\omega, \mu)}{\omega} \,
J_{\perp}(n \cdot p, 0, \omega, \mu) \nonumber \\
&& \hspace{1.0 cm} + \int_0^{\omega_s} \,\,d \omega^{\prime} \,\,  \left [ \frac{n \cdot p}{m_{\rho}^2} \,
{\rm Exp} \left [{m_{\rho}^2 - \omega^{\prime} \, n \cdot p \over n \cdot p \, \omega_M} \right ]
- {1 \over \omega^{\prime}} \right ] \, \phi_{B, {\rm eff}}^{+}(\omega^{\prime},\mu) \, \bigg \}  \,, \nonumber \\
&& \equiv F_{V, 2P}^{\rm {LP}}(n \cdot p) + F_{V, 2P}^{\rm{NLP}}(n \cdot p) \,,
\label{NLL 2-particle contribution to form factors}
\end{eqnarray}
where $F_{V, 2P}^{\rm LP}$ and $F_{V, 2P}^{\rm NLP}$ are defined by keeping only the first and the second terms
in the bracket, respectively.
In addition, the effective ``distribution amplitude" $\phi_{B, {\rm eff}}^{+}(\omega^{\prime},\mu)$
\begin{eqnarray}
\phi_{B, {\rm eff}}^{+}(\omega^{\prime},\mu)
&=& \phi_{B}^{+}(\omega^{\prime},\mu) +{\alpha_s(\mu) \, C_F \over 4 \,  \pi} \,
\bigg \{ \int_0^{\omega^{\prime}} \,\,d \omega  \, \left [ {2 \over \omega-\omega^{\prime}} \,
\ln { \mu^2 \over n \cdot p \, (\omega^{\prime}-\omega)}  \right ]_{\oplus} \,
\phi_{B}^{+}(\omega,\mu)  \nonumber \\
&&  - \, \omega^{\prime} \, \int_0^{\omega^{\prime}} \,\,d \omega   \,
\left [ {1 \over \omega-\omega^{\prime}} \,
\ln { \omega^{\prime} -\omega \over \omega}  \right ]_{\oplus} \,
{\phi_{B}^{+}(\omega^{\prime},\mu) \over \omega}  \nonumber \\
&& + \, {\omega^{\prime} \over 2} \, \int_0^{\infty} \,\,d \omega   \,
\left [  \ln^2 \bigg|{ \omega-\omega^{\prime} \over \omega^{\prime}} \bigg| \, \right ]\,
{d \over d \omega} \, {\phi_{B}^{+}(\omega,\mu) \over \omega}  \, \nonumber \\
&& - \, \int_{\omega^{\prime}}^{\infty} \,\,d \omega   \,
\left [ \ln { \mu^2 \over n \cdot p \, \omega^{\prime}} - {\pi^2 \over 2} - 1 \right ] \,
{d \over d \omega} \, \phi_{B}^{+}(\omega,\mu)  \, \nonumber \\
&& + \, \omega^{\prime} \, \int_{\omega^{\prime}}^{\infty} \,\,d \omega  \,
\bigg [ \ln { \mu^2 \over n \cdot p \, \omega^{\prime}} \,
\ln { \omega -\omega^{\prime} \over \omega^{\prime}}
- {1 \over 2} \, \ln^2 { \mu^2 \over n \cdot p \, (\omega-\omega^{\prime})}
+  {1 \over 2} \, \ln^2 { \mu^2 \over n \cdot p \, \omega^{\prime} }  \nonumber \\
&& + \, 3 \, \ln { \omega -\omega^{\prime} \over \omega^{\prime}} -{2 \, \pi^2 \over 3}  \bigg ]  \, \,
{d \over d \omega} \, {\phi_{B}^{+}(\omega,\mu) \over \omega}  \,  \bigg \} \,.
\label{effective B meson DA}
\end{eqnarray}
is introduced to absorb the next-to-leading-order (NLO) hard-collinear correction to the general $B \to \gamma^{\ast}$
transition form factors.
It is evident that $F_{V, 2P}^{\rm LP}$ corresponds to the leading-power contribution to
the on-shell $B \to \gamma$ form factors computed from
QCD factorization and the corresponding convolution integral can be expressed in terms of the moments of
the $B$-meson DA as \cite{Beneke:2011nf}:
\begin{eqnarray}
&& \int_0^{\infty} \, d \omega \, \frac{\phi_B^{+}(\omega, \mu)}{\omega} \,  \,
J_{\perp}(n \cdot p, 0, \omega, \mu)  \nonumber \\
&& = \lambda_B^{-1}(\mu) \, \bigg \{  1 + {\alpha_s(\mu) \, C_F \over 4 \,  \pi} \,
\bigg [\sigma_2(\mu) + 2 \, \ln {\mu^2 \over n \cdot p \, \mu_0} \, \sigma_1(\mu)
+ \ln^2 {\mu^2 \over n \cdot p \, \mu_0} - {\pi^2 \over 6} -1 \bigg ] \bigg \}  \,.
\label{QCDF for B to gamma FFs}
\end{eqnarray}
Here, a hadronic reference scale $\mu_0= 1 \, {\rm GeV}$ is introduced in the definition
of the inverse-logarithmic moments, in contrast to \cite{Braun:2003wx},
 to avoid the appearance of a parametrically large logarithm
due to the scale evolution of $\sigma_n(\mu)$ \cite{Beneke:2011nf}.

Several comments on the nonperturbative modification of the spectral function shown in
the second term in the  bracket of (\ref{NLL 2-particle contribution to form factors})
are in order.
\begin{itemize}
\item{In light of the power counting rule (\ref{power counting for SR parameters})
and the scaling $\omega \sim \Lambda$ due to the canonical behaviour of the $B$-meson DA $\phi_B^{+}(\omega,\mu)$,
one can readily identify that the logarithmic terms $\ln^2 \left|{ \omega-\omega^{\prime} \over \omega^{\prime}} \right|$
and $\ln \left ({\omega-\omega^{\prime} \over \omega^{\prime}} \right )$ involved in (\ref{effective B meson DA}) need to be
counted as $\ln^2 \left ( {n \cdot p \over \Lambda} \right )$ and $\ln \left ( {n \cdot p \over \Lambda} \right )$
in the heavy quark limit. The appearance of such large logarithms can be traced back to the continuum subtraction
in the construction of QCD sum rules for the end-point contribution to the $B \to \gamma$ form factors, with the aid of
the parton-hadron duality approximation. This observation appears to indicate that the ``hadronic" photon  contribution
to the $B \to \gamma \ell \nu$ amplitude suffers from rapidity divergences in QCD factorization, which are regularized
by the nonperturbative parameter $\omega_M$ in the sum rule approach.}
\item{In the absence of a detailed analysis of the subleading form factor $\xi(E_\gamma)$
in QCD \cite{Beneke:2011nf}, the precise relation between the end-point contribution
computed from the hadronic dispersion relations and $\xi(E_\gamma)$ cannot be established in a model-independent way.
It is rather plausible that adding up the symmetry-conserving form factor $\xi(E_\gamma)$ and the soft two-particle
correction together would  yield double counting of quark-gluon and hadron degrees of freedom.}
\end{itemize}

\section{Three-particle subleading power contribution}
\label{sect:three-particle contribution}

\begin{figure}[t]
\begin{center}
\includegraphics[width=0.30 \textwidth]{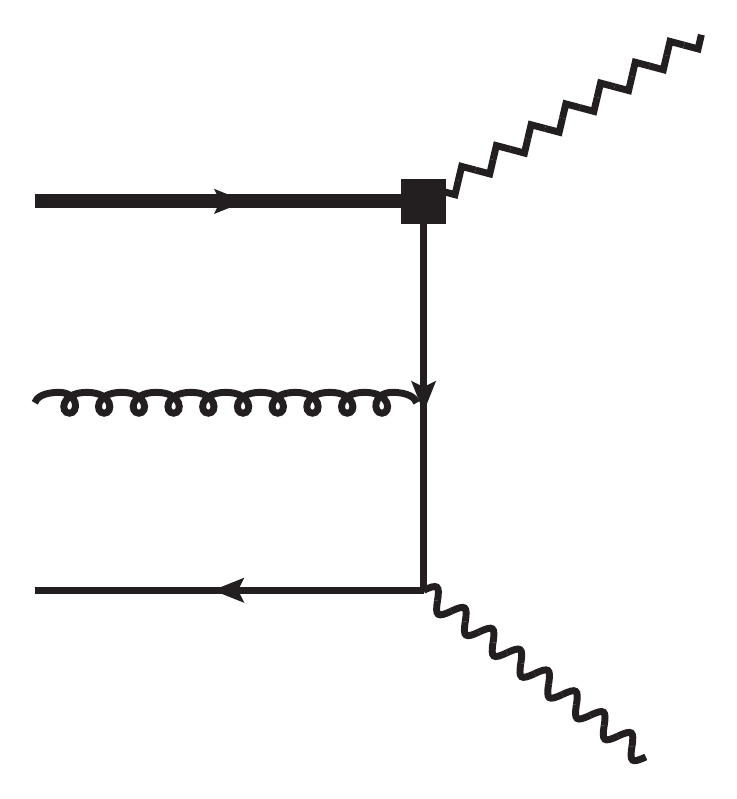}
\end{center}
\caption{Higher-twist contribution to the correlation function (\ref{def: correlation function})
from the three-particle DAs of the $B$-meson at tree level.
Same conventions as in figure \ref{tree diagram of the correlator}. }
\label{3-particle diagram of the correlator}
\end{figure}

The purpose of this section is to compute the higher twist contributions to the on-shell
$B \to \gamma$ transition form factors from the three-particle $B$-meson  DAs at tree level.
Following the main idea of the dispersion approach discussed in Section \ref{sect: dispersion approach},
we need to establish the factorization formula for the three-particle contribution to the generalized
$B \to \gamma^{\ast}$ form factors at leading order in $\alpha_s$.
This amounts to evaluating the subleading power contribution (compared with the two-particle contribution
shown in (\ref{resummation improved factorization formula}))  induced by the partonic diagram
displayed in figure \ref{3-particle diagram of the correlator}.

Applying the light-cone expansion of the light-quark propagator in the background gluon field
\cite{Balitsky:1987bk} and keeping  the one-gluon part without the covariant derivative of the $G_{\mu \nu}$ terms
\begin{eqnarray}
\langle 0 | T \{ q(x) \,, \bar q(0) \} | 0 \rangle |_{G}
\supset  - {i \over 16 \, \pi^2}  \, {1 \over x^2} \,
\int_0^1 d u \, \left [ \! \not x \, \sigma_{\alpha \beta}
- 4 \, i \, u \, x_{\alpha} \, \gamma_{\beta} \right ] \, G^{\alpha \beta}(u \, x)  \,,
\end{eqnarray}
one can readily obtain the factorization formula for the
three-particle contribution to the form factors $F_{V}^{B \to \gamma^{\ast}}$
and $\hat{F}_{A}^{B \to \gamma^{\ast}}$ at tree level
\begin{eqnarray}
&& F_{V, \, 3P}^{B \to \gamma^{\ast}}(n \cdot p, \bar n \cdot p)
= \hat{F}_{A, \, 3P}^{B \to \gamma^{\ast}}(n \cdot p, \bar n \cdot p)  \nonumber  \\
&& = - \frac{Q_u \, \tilde{f}_B(\mu) \, m_B}{(n \cdot p)^2} \,
\int_0^{\infty} d \omega \, \int_0^{\infty} d \xi \, \int_0^1 d u \,
\bigg \{\frac{ \rho_{3P}^{(2)}(u,\omega,\xi)}{[\bar n \cdot p - \omega - u \, \xi]^2}
+ \frac{\rho_{3P}^{(3)}(u,\omega,\xi)}{[\bar n \cdot p - \omega - u \, \xi]^3} \bigg \} \,,
\label{3-particle contribution to B to gamma-star FFs}
\end{eqnarray}
by employing the following Fourier integral in Minkowski space
\begin{eqnarray}
\int d^4 x \, e^{i q \cdot x} \,\, {x_{\mu} \over x^2}
= \frac{8 \,\pi^2}{q^4} \, q_{\mu} \,.
\end{eqnarray}
The explicit expressions for $\rho_{3P}^{(i)} \, (i=2, 3)$  are given by
\begin{eqnarray}
&& \rho_{3P}^{(2)}(u,\omega,\xi) = \Psi_V(\omega, \xi) + (1 + 2 \, u) \, \Psi_A(\omega, \xi)  \,,  \qquad
\rho_{3P}^{(3)}(u,\omega,\xi) = -2 \,  (1 + 2 \, u) \, \bar X_A(\omega, \xi)  \,,  \nonumber \\
&& \bar X_A(\omega, \xi) = \int_0^{\omega} \, d \eta \, X_A(\eta, \xi)  \,, \qquad \hspace{3.0 cm}
\bar Y_A(\omega, \xi) = \int_0^{\omega} \, d \eta \, Y_A(\eta, \xi)   \,,
\end{eqnarray}
where the relevant three-particle DAs of the $B$-meson are defined by
the following light-cone  matrix element \cite{Kawamura:2001jm,Geyer:2005fb}
\begin{eqnarray}
&& \langle 0 |\bar u_{\alpha}(x) \,\, G_{\lambda \, \rho}(u \, x)
 \,\, b_{v \beta}(0)| B^{-}(v)\rangle  \big |_{x^2=0} \nonumber \\
&& = {\tilde{f}_B(\mu) \, m_B \over 4} \, \int_0^{\infty} d \omega \, \int_0^{\infty} d \xi \,\,
e^{-i(\omega+ u\, \xi) \, v \cdot x} \,\,
\bigg [ \left (1 + \! \not v \right ) \,
\bigg \{  (v_{\lambda} \, \gamma_{\rho} - v_{\rho} \, \gamma_{\lambda} ) \,
\left [\Psi_A(\omega, \xi) - \Psi_V(\omega, \xi) \right ]  \nonumber \\
&& \hspace{0.6 cm} - i \, \sigma_{\lambda \rho} \, \Psi_V(\omega, \xi)
- \frac{x_{\lambda} v_{\rho} - x_{\rho} v_{\lambda} }{v \cdot x} \, X_A(\omega, \xi)
+ \frac{x_{\lambda} \gamma_{\rho} - x_{\rho} \gamma_{\lambda} }{v \cdot x} \, Y_A(\omega, \xi)
\bigg \} \, \gamma_5 \bigg ]_{\beta \alpha} \,,
\end{eqnarray}
with the soft Wilson lines on the left-hand side omitted for brevity.
The three-particle DAs $\Psi_V$, $\Psi_A$, $X_A$ and $Y_A$ depend on two light-cone variables
$\omega=\bar n \cdot k_1$ and $\xi=\bar n \cdot k_2$, where $k_1$ and $k_2$ are the light-quark and gluon momenta
inside the $B$-meson. In contrast to the two-particle $B$-meson DAs, model-independent properties
of the three-particle DAs in QCD, including the RG evolution equations and the asymptotic behaviours
for $\omega, \xi \gg \Lambda$, are poorly explored at present (see \cite{Braun:2015pha} for an exception).

The tree-level factorization formula for the three-particle contribution to the on-shell $B \to \gamma$
form factors can be readily constructed  by setting $\bar n \cdot p \to 0$
in  (\ref{3-particle contribution to B to gamma-star FFs}):
\begin{eqnarray}
&& F_{V, \, 3P}^{B \to \gamma}(n \cdot p)
= \hat{F}_{A, \, 3P}^{B \to \gamma}(n \cdot p)  \nonumber  \\
&& = - \frac{Q_u \, \tilde{f}_B(\mu) \, m_B}{(n \cdot p)^2} \,
\int_0^{\infty} d \omega \, \int_0^{\infty} d \xi \, \int_0^1 d u \,
\bigg \{\frac{ \rho_{3P}^{(2)}(u,\omega,\xi)}{[\omega + u \, \xi]^2}
- \frac{\rho_{3P}^{(3)}(u,\omega,\xi)}{[\omega + u \, \xi]^3} \bigg \} \, \nonumber \\
&& =  - \frac{Q_u \, \tilde{f}_B(\mu) \, m_B}{(n \cdot p)^2} \,
\int_0^{\infty} d \omega \, \int_0^{\infty} d \xi \,
\bigg \{ \frac{1}{\omega(\omega+\xi)} \, \Psi_V(\omega, \xi)  \nonumber \\
&&  \hspace{0.5 cm} +  \left [ \frac{1}{\omega(\omega+\xi)}  - \frac{2}{\xi(\omega+\xi)}
+{2 \over \xi^2} \, \ln {\omega+\xi \over \omega} \right ]  \,\, \Psi_A(\omega, \xi)
+ { 4 \, \omega + \xi \over \omega^2 \, (\omega+\xi)^2 }  \, \bar X_A(\omega, \xi)   \bigg \} \,.
\label{3-particle contribution to B to gamma FFs}
\end{eqnarray}
In light of the end-point behaviours of the three-particle $B$-meson DAs at $\omega, \, \xi \to 0$
from a QCD sum rule analysis \cite{Khodjamirian:2006st}
\begin{eqnarray}
\Psi_V(\omega, \xi) \sim \Psi_A(\omega, \xi) \sim \xi^2 \,,  \qquad
X_A(\omega, \xi) \sim \xi (2 \omega - \xi) \,, \qquad Y_A(\omega, \xi) \sim \xi \,,
\end{eqnarray}
it is straightforward to verify that the convolution integral of the short-distance function
with the $B$-meson DAs in (\ref{3-particle contribution to B to gamma FFs})
suffers from rapidity divergences as speculated in \cite{Braun:2012kp}.
We therefore conclude that decomposing the three-particle contribution
of the on-shell $B \to \gamma$ form factors into the factorizable  effect computed from light-cone OPE
and the nonperturbative modification as displayed in
(\ref{modified master formula of FV}) and (\ref{modified master formula of FAhat}) cannot be justified,
and instead one needs to employ the original dispersion expressions
presented in (\ref{master formula of FV}) and (\ref{master formula of FAhat}).

Extracting the spectral function of (\ref{3-particle contribution to B to gamma-star FFs}) in the
variable $\bar n \cdot p$ and substituting it into  (\ref{master formula of FV}) and (\ref{master formula of FAhat})
give rise to the desired three-particle contribution to the $B \to \gamma$ form factors
\begin{eqnarray}
&& F_{V, \, 3P}(n \cdot p)
= \hat{F}_{A, \, 3P}(n \cdot p)  \nonumber  \\
&& = - \frac{Q_u \, \tilde{f}_B(\mu) \, m_B}{(n \cdot p)^2} \,
\left \{  \frac{n \cdot p}{m_{\rho}^2} \,
{\rm Exp} \left [{m_{\rho}^2 \, \over n \cdot p \, \omega_M} \right ] \,
I_{3P}^{\rm I}(\omega_s, \omega_M) +  I_{3P}^{\rm II}(\omega_s, \omega_M) \right \} \,,
\label{3-particle contribution to B to gamma FFs: dispersion approach}
\end{eqnarray}
where the coefficient functions entering (\ref{3-particle contribution to B to gamma FFs: dispersion approach}) are
\begin{eqnarray}
&& I_{3P}^{\rm I}(\omega_s, \omega_M) \nonumber \\
&& = \int_0^{\omega_s} \, d \omega \, \int_{\omega_s-\omega}^{\infty} \,
{d \xi \over \xi} \, e^{-\omega_s/\omega_M} \, \left [ \rho_{3P}^{(2)}(u,\omega,\xi)
- {1 \over 2} \, {d \over d \omega} \,  \rho_{3P}^{(3)}(u,\omega,\xi)
- { \rho_{3P}^{(3)}(u,\omega,\xi)  \over 2 \, \omega_M} \right ] \Bigg |_{u= {\omega_s-\omega \over \xi}} \nonumber \\
&& \hspace{0.5 cm} + \int_0^{\omega_s} \, d \omega^{\prime} \, \int_0^{\omega^{\prime}} \, d \omega \,
\int_{\omega^{\prime} -\omega}^{\infty} \, {d \xi \over \xi} \, e^{-\omega^{\prime}/\omega_M} \,
{1 \over \omega_M} \, \left [\rho_{3P}^{(2)}(u,\omega,\xi) - {\rho_{3P}^{(3)}(u,\omega,\xi)  \over 2 \, \omega_M} \,
\right ]\Bigg |_{u= {\omega^{\prime}-\omega \over \xi}}  \,, \\
&& I_{3P}^{\rm II}(\omega_s, \omega_M) \nonumber \\
&&= - \int_0^{\omega_s} \, d \omega \, \int_{\omega_s-\omega}^{\infty} \,
{d \xi \over \xi} \, {1 \over \omega_s} \, \left [ \rho_{3P}^{(2)}(u,\omega,\xi)
- {1 \over 2} \, {d \over d \omega} \,  \rho_{3P}^{(3)}(u,\omega,\xi)
- { \rho_{3P}^{(3)}(u,\omega,\xi)  \over 2 \, \omega_M} \right ] \Bigg |_{u= {\omega_s-\omega \over \xi}} \nonumber \\
&& \hspace{0.5 cm} + \int_{\omega_s}^{\infty} \, d \omega^{\prime} \, \int_0^{\omega^{\prime}} \, d \omega \,
\int_{\omega^{\prime} -\omega}^{\infty} \, {d \xi \over \xi} \, {1 \over (\omega^{\prime})^2} \,
\left [\rho_{3P}^{(2)}(u,\omega,\xi) - {\rho_{3P}^{(3)}(u,\omega,\xi)  \over 2 \, \omega_M} \,
\right ]\Bigg |_{u= {\omega^{\prime}-\omega \over \xi}}  \,.
\end{eqnarray}
Employing the canonical behaviours of the three-particle $B$-meson DAs and the
power counting rule (\ref{power counting for SR parameters}) for the sum rule parameters leads to
\begin{eqnarray}
I_{3P}^{\rm I}(\omega_s, \omega_M) \sim {\cal O}(\Lambda^2 /m_b) \,, \qquad
I_{3P}^{\rm II}(\omega_s, \omega_M) \sim {\cal O}(1)  \,,
\end{eqnarray}
which implies that both the ``hard" and ``soft" contributions to the $B \to \gamma$ form factors from
the three-particle $B$-meson DAs scale as $(\Lambda/m_b)^{3/2}$ in the heavy quark limit,
in contrast to the two-particle ``hard" and ``soft" contributions discussed before.
Such observation can be also inferred from the violation of QCD factorization for the three-particle
contribution to the form factors $F_V(n \cdot p)$  and $\hat{F}_A(n \cdot p)$,
 due to the rapidity divergences, indicating that the intuitive correspondence between the power expansion
 and the dynamical twist expansion can be spoiled by the soft corrections \cite{Agaev:2010aq}.

Adding up different pieces together, we obtain the final expressions for the on-shell $B \to \gamma$ form factors
in the dispersion approach
\begin{eqnarray}
F_V(n \cdot p) = F_{V, 2P}(n \cdot p) + F_{V, 3P}(n \cdot p)  + F_{V, NLP}^{\rm LC}(n \cdot p) \,,
\label{final expression of FV}  \\
\hat{F}_A(n \cdot p) = \hat{F}_{A, 2P}(n \cdot p) + \hat{F}_{A, 3P}(n \cdot p)
+ \hat{F}_{A, NLP}^{\rm LC}(n \cdot p) \,,
\label{final expression of FAhat}
\end{eqnarray}
where the manifest expressions of  individual terms on the right-hand side of (\ref{final expression of FV})
and (\ref{final expression of FAhat}) are given by (\ref{NLL 2-particle contribution to form factors}),
(\ref{3-particle contribution to B to gamma FFs: dispersion approach}) and (\ref{subleading power local contribution}),
respectively. The following comments on the structures of the form factors $F_V(n \cdot p)$ and $\hat{F}_A(n \cdot p)$
displayed in (\ref{final expression of FV}) and  (\ref{final expression of FAhat}) can be made.
\begin{itemize}
\item{The symmetry-violating contribution to the on-shell $B \to \gamma$ form factors comes solely from
the local subleading power  corrections as indicated by $F_{V, NLP}^{\rm LC}$ and $\hat{F}_{A, NLP}^{\rm LC}$.
The non-local subleading power contributions from the end-point region preserve the symmetry relation
of $F_V$ and $\hat{F}_A$ due to the helicity conservation, in support of a similar observation made in \cite{Beneke:2011nf}
applying the QCD factorization approach.}
\item {Despite of the fact that the leading-power contribution to the generalized $B \to \gamma^{\ast}$ form factors
originates from the two-particle $B$-meson DA $\phi_B^{+}(\omega, \mu)$,  the end-point (``soft") contributions
to the on-shell $B \to \gamma$ form factors from both the two-particle and three-particle DAs contribute at the same power
in the heavy-quark expansion. Following the arguments in \cite{Agaev:2010aq}, yet higher-twist corrections
from the four-particle $B$-meson DAs would also generate the subleading power contribution suppressed
by one power of $\Lambda/m_b$, when compared with the leading-twist contribution.
We will leave a transparent demonstration of this interesting pattern for a future work, by including
the two-gluon field strength terms and the covariant derivative of the $G^{\mu \nu}$ terms in the light-cone
expansion of the massless-quark propagator in the background gluon field. }
\end{itemize}

\section{Numerical analysis}
\label{sect:Numerical analysis}

We are now in a position to explore the phenomenological consequence of the subleading power corrections
to the $B \to \gamma$ form factors computed from the dispersion approach.
In order to perform the numerical analysis of the newly derived expressions for $F_V(n \cdot p) $
and $\hat{F}_A(n \cdot p) $ in  (\ref{final expression of FV}) and  (\ref{final expression of FAhat}),
we will proceed by specifying the nonperturbative models for the two-particle and three-particle DAs of the $B$-meson,
determining the sum rule parameters and setting the hard and hard-collinear scales.
Taking advantage of  the new measurements of the partial branching fractions of $B \to \gamma \ell  \nu$
from the Belle Collaboration \cite{Heller:2015vvm}, theory constraints of the inverse moment of
the leading-twist DA $\phi_B^{+}(\omega, \mu)$ will be further addressed with
the updated predictions for the $B \to \gamma$ form factors presented above.

\subsection{Theory input parameters}

Following \cite{Braun:2012kp}, we will consider two models of the two-particle $B$-meson DA
$\phi_B^{+}(\omega, \mu_0)$ motivated from the QCD sum rule analysis at tree level \cite{Grozin:1996pq}
and at NLO  \cite{Braun:2003wx}
\begin{eqnarray}
&&  \phi_{B,\rm I}^+(\omega,\mu_0) = \frac{\omega}{\omega_0^2} \, e^{-\omega/\omega_0} \,,
\label{the first model of the B-meson DA} \\
&&  \phi_{B,\rm II}^+(\omega,\mu_0)= \frac{1}{4 \pi \,\omega_0} \, {k \over k^2+1} \,
\left[ {1 \over k^2+1} - \frac{2 (\sigma_{1}(\mu_0) -1)}{\pi^2}  \, \ln k \right ] \,, \hspace{0.5 cm}
k= \frac{\omega}{1 \,\, \rm GeV} \,,
\label{the second model of the B-meson DA}
\end{eqnarray}
where the shape parameter $\omega_0=\lambda_B(\mu_0)$.
As emphasized in \cite{Wang:2015vgv}, the above models can only serve as a reasonable description of
$\phi_{B}^+(\omega,\mu_0)$ at small $\omega$ and they could not reproduce the model-independent behaviour at
large $\omega$ predicted from perturbative QCD. Since the dominant contributions to the $B \to \gamma$ form factors
come from the small $\omega$ region according to the power counting analysis, we will not improve the above models
for the $B$-meson DA $\phi_{B}^+(\omega,\mu_0)$ by implementing the perturbative constraints as discussed in \cite{Feldmann:2014ika}.
In particular, the leading power contribution to the on-shell $B \to \gamma$ form factors
is insensitive to precise shape of $\phi_{B}^+(\omega,\mu_0)$ at small $\omega$, and
only depends on the inverse-logarithmic moments as shown in (\ref{QCDF for B to gamma FFs}).
Applying the one-loop evolution equation of $\phi_{B}^+(\omega,\mu)$ in (\ref{Lange-Neubert equation})
leads to \cite{Beneke:2011nf}
\begin{eqnarray}
\frac{\lambda_B(\mu_0)}{\lambda_B(\mu)} &=&
1 + {\alpha_s(\mu_0) \, C_F \over 4 \, \pi} \, \ln {\mu \over \mu_0} \,
\left [2 - 2\, \ln {\mu \over \mu_0} - 4 \, \sigma_{1}(\mu_0) \right ] + {\cal O}(\alpha_s^2)\,.
\label{lambdab evolution}
\end{eqnarray}
For the inverse-logarithmic  moments $\sigma_1$ and $\sigma_2$, we will take
$\sigma_1(\mu_0)=1.5 \pm 1$ and $\sigma_2(\mu_0)=3 \pm 2$ from \cite{Beneke:2011nf},
and the scale evolution effect of these parameters is not needed for the evaluation of the leading power
contribution to the $B \to \gamma$ form factors at NLL.

For the three-particle $B$-meson DAs, we adopt an exponential model in consistent with the small
$\omega, \xi$ behaviour from the tree-level QCD sum rule analysis \cite{Khodjamirian:2006st}
\begin{eqnarray}
\Psi_V(\omega, \xi, \mu_0) &=& \Psi_A(\omega, \xi, \mu_0)
=\frac{\lambda_E^2}{6 \, \omega_0^4} \, \xi^2 \, e^{-(\omega+\xi)/\omega_0} \,, \nonumber \\
X_A(\omega, \xi, \mu_0) &=& \frac{\lambda_E^2}{6 \, \omega_0^4} \, \xi \, (2 \, \omega - \xi) \,
e^{-(\omega+\xi)/\omega_0} \,, \nonumber \\
Y_A(\omega, \xi, \mu_0) &=& - \frac{\lambda_E^2}{24 \, \omega_0^4} \, \xi \,
(7 \, \omega_0 - 13 \, \omega + 3 \, \xi) \, e^{-(\omega+\xi)/\omega_0} \,,
\end{eqnarray}
where the normalization parameter computed from QCD sum rules including the higher-order
perturbative and nonperturbative effects is determined to be
$\lambda_E^2(\mu_0)=(0.03 \pm 0.02 ) \, {\rm GeV^2}$ \cite{Nishikawa:2011qk}.
It needs to point out that we neglect the small correction
due to the difference $(\Psi_A  - \Psi_V) \sim (\lambda_E^2-\lambda_H^2) \, \omega \, \xi^2$ which can be extracted
from the NLO QCD correction to the sum rules for the three-particle DAs
derived in \cite{Khodjamirian:2006st}, and the normalization coefficients in front of the DAs $X_A$ and $Y_A$
can also differ from $\lambda_E^2$ in general.

Now we turn to the determination of the Borel parameter $\omega_M$  and the duality-threshold parameter $\omega_s$
entering the expressions for $F_{V, 2P}$ and $F_{V, 3P}$. The general procedure to
choose the sum rule parameters satisfying with the power counting rule (\ref{power counting for SR parameters})
has been discussed in \cite{Wang:2015vgv}, and repeating the same strategies gives rise to the following intervals
\begin{eqnarray}
M^2 \equiv n \cdot p \,\, \omega_M = (1.25 \pm 0.25) \, {\rm GeV^2} \,, \qquad
s_0  \equiv n \cdot p \,\, \omega_s = (1.50 \pm 0.20) \, {\rm GeV^2}  \,,
\end{eqnarray}
in agreement with the values used for the LCSR for
 the $\gamma^{\ast} \to \pi \gamma$ form factor \cite{Agaev:2010aq}. Note that the effective threshold
$\omega_s$ in the dispersion expressions for the $B \to \gamma$ form factors should be compared to
that adopted in the two-point sum rules for the $\rho$-meson channel \cite{Khodjamirian:2006st}.

The HQET decay constant of the $B$-meson $\tilde{f}_B(\mu)$ will be traded into the QCD decay constant
$f_B$ with the matching equation (\ref{matching condition for the fB}), which will be computed from the
two-point QCD sum rules including ${\cal O}(\alpha_s)$ corrections to the perturbative contribution and
the quark-gluon condensate operator contributions up to dimension-6 \cite{Duplancic:2008ix}.
We will take the same intervals of the Borel mass and the threshold parameter
\begin{eqnarray}
\overline{M}^2=(5.0 \pm 1.0) \, {\rm GeV^2} \,, \qquad
\bar{s}_0=(35.6^{+2.1}_{-0.9})  \, {\rm GeV^2} \,
\end{eqnarray}
as adopted in \cite{Duplancic:2008ix,Wang:2015vgv}.
For the hard scales involved in the hard matching coefficients, we will choose
${m_b / 2} \leq \mu_{h1}=\mu_{h2} \leq 2 \, m_b$ with the default
value  $\mu_{h1}=\mu_{h2}=m_b$.
The factorization scale $\mu$ will be varied in the interval
$1 \, {\rm GeV} \leq \mu \leq 2 \, {\rm GeV}$ around the central value $1.5 \, {\rm GeV}$.
Furthermore, we will use the values of the bottom-quark mass in the ${\rm\overline{MS}}$ scheme
$\overline{m_b}(\overline{m_b})= (4.193^{+0.022}_{-0.035}) \, {\rm GeV}$
determined from non-relativistic sum rules \cite{Beneke:2014pta}.

\subsection{Predictions for the $B \to \gamma$ form factors}

\begin{figure}
\begin{center}
\includegraphics[width=0.80 \textwidth]{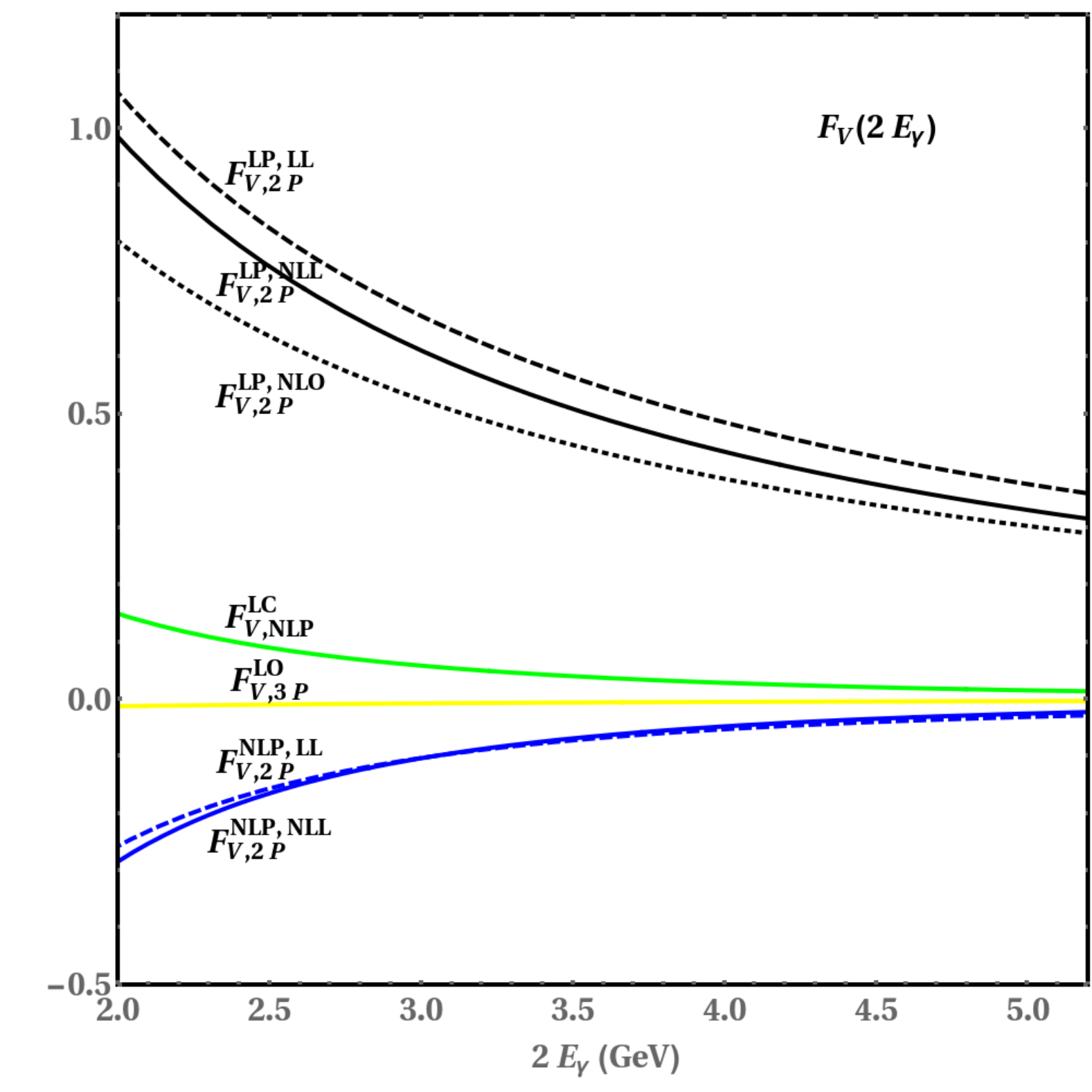}
\end{center}
\caption{The photon-energy dependence of various contributions to the form factor $F_V(2 \, E_{\gamma})$,
with the exponential model of $\phi_B^{+}(\omega, \mu_0)$ and
the inverse moment $\lambda_B(\mu_0)=354 \, {\rm MeV}$ determined in \cite{Wang:2015vgv}.
The separate contributions correspond to the leading power two-particle effect at leading logarithmic (LL)
accuracy ($F_{V, 2P}^{\rm{LP, LL}}$, dashed black), at NLO ($F_{V, 2P}^{\rm{LP, NLO}}$, dotted black),
and at NLL ($F_{V, 2P}^{\rm{LP, NLL}}$, solid black);  the subleading power two-particle correction
at LL ($F_{V, 2P}^{\rm{NLP, LL}}$, dashed blue), and at NLL ($F_{V, 2P}^{\rm{NLP, NLL}}$, solid blue);
the subleading power three-particle correction at LO  ($F_{V, 3P}^{\rm{LO}}$, solid yellow);
and the power suppressed local effect at tree level ($F_{V, NLP}^{\rm{LC}}$, solid green).}
\label{Breakdown of different contributions for the energy dependence}
\end{figure}

Now we are ready to investigate the numerical impact of the subleading power contributions
from the two-particle and three-particle $B$-meson DAs on the  $B \to \gamma$  form factors.
To develop a transparent understanding of the newly calculated corrections in this work,
we display the photon-energy dependence of the leading power  two-particle contribution,
the subleading power two-particle and three-particle corrections as well as the power suppressed
local contribution in figure \ref{Breakdown of different contributions for the energy dependence},
where we take $\phi_{B,\rm I}^+(\omega,\mu_0) $ as a default model with
$\lambda_B(\mu_0)=354^{+38}_{-30} \, {\rm MeV}$ determined from \cite{Wang:2015vgv}.
One can readily find that, with the adopted value of $\lambda_B(\mu_0)$,
the subleading power two particle contribution  $F_{V, 2P}^{\rm{NLP, NLL}}( n \cdot p)$
including the NLL resummation effect can decrease the leading-power prediction for the form factor $F_V(n \cdot p)$
by approximately $(10 \sim 30) \%$ in the kinematic region $ n \cdot p \in  \left [ 2 \, {\rm GeV} \,, m_B \right ]$;
while the power suppressed correction from the three-particle $B$-meson DAs at tree level only induce a minor impact on the
theory prediction of $F_V(n \cdot p)$ and  numerically ${\cal O}(1 \, \%)$.
We also find that perturbative QCD corrections to the ``soft" two-particle contribution can shift the tree-level
prediction  $F_{V, 2P}^{\rm{NLP, LL}}$ by an amount of $(10 \sim 20) \%$ with the default theory inputs.
We are therefore led to conclude that the power suppressed corrections to the $B \to \gamma$ form factors
are dominated by the soft two-particle contribution at tree level with the default model of $B$-meson DAs.

\begin{figure}
\begin{center}
\includegraphics[width=0.45 \textwidth]{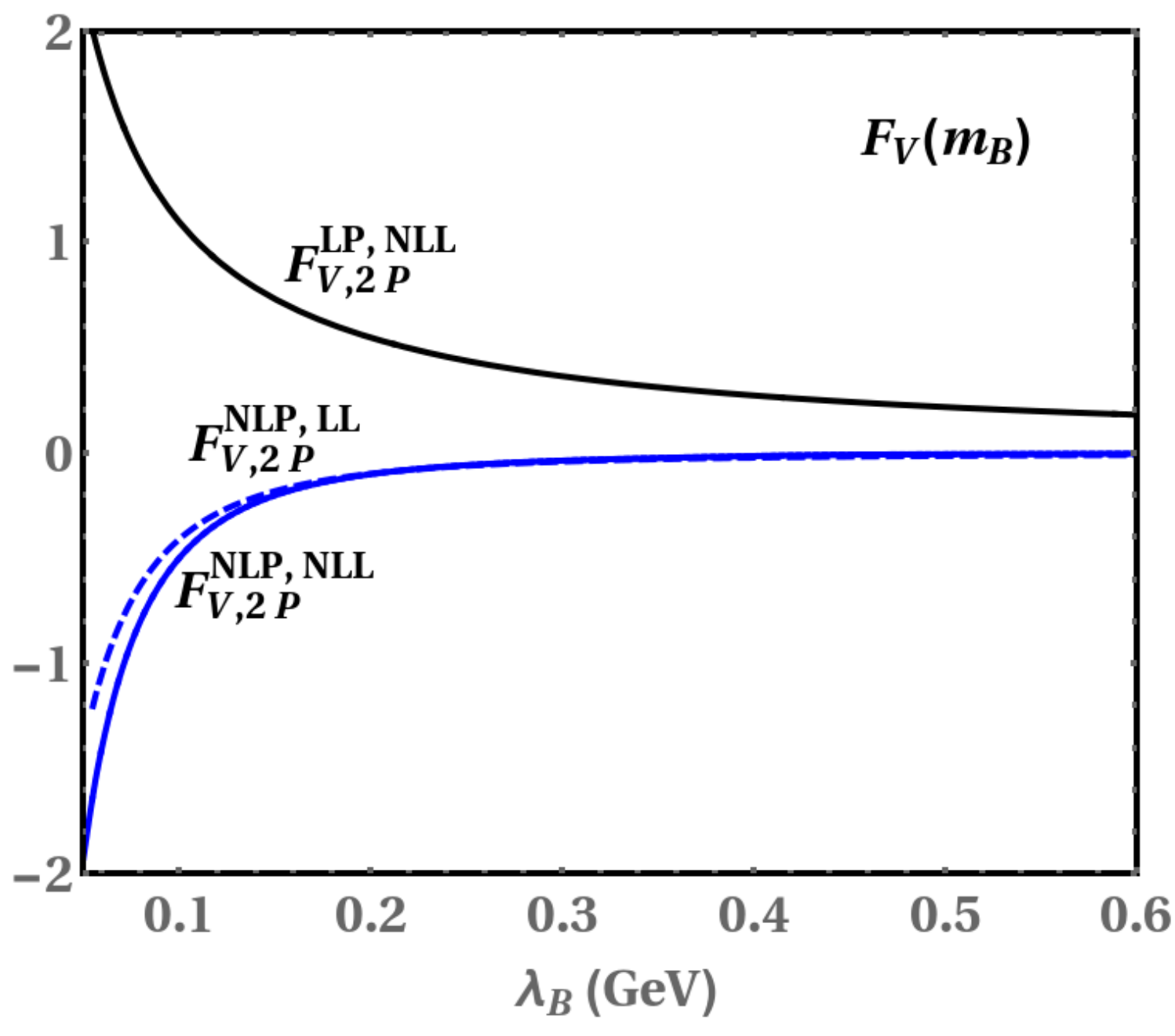} \hspace{0.5 cm}
\includegraphics[width=0.45 \textwidth]{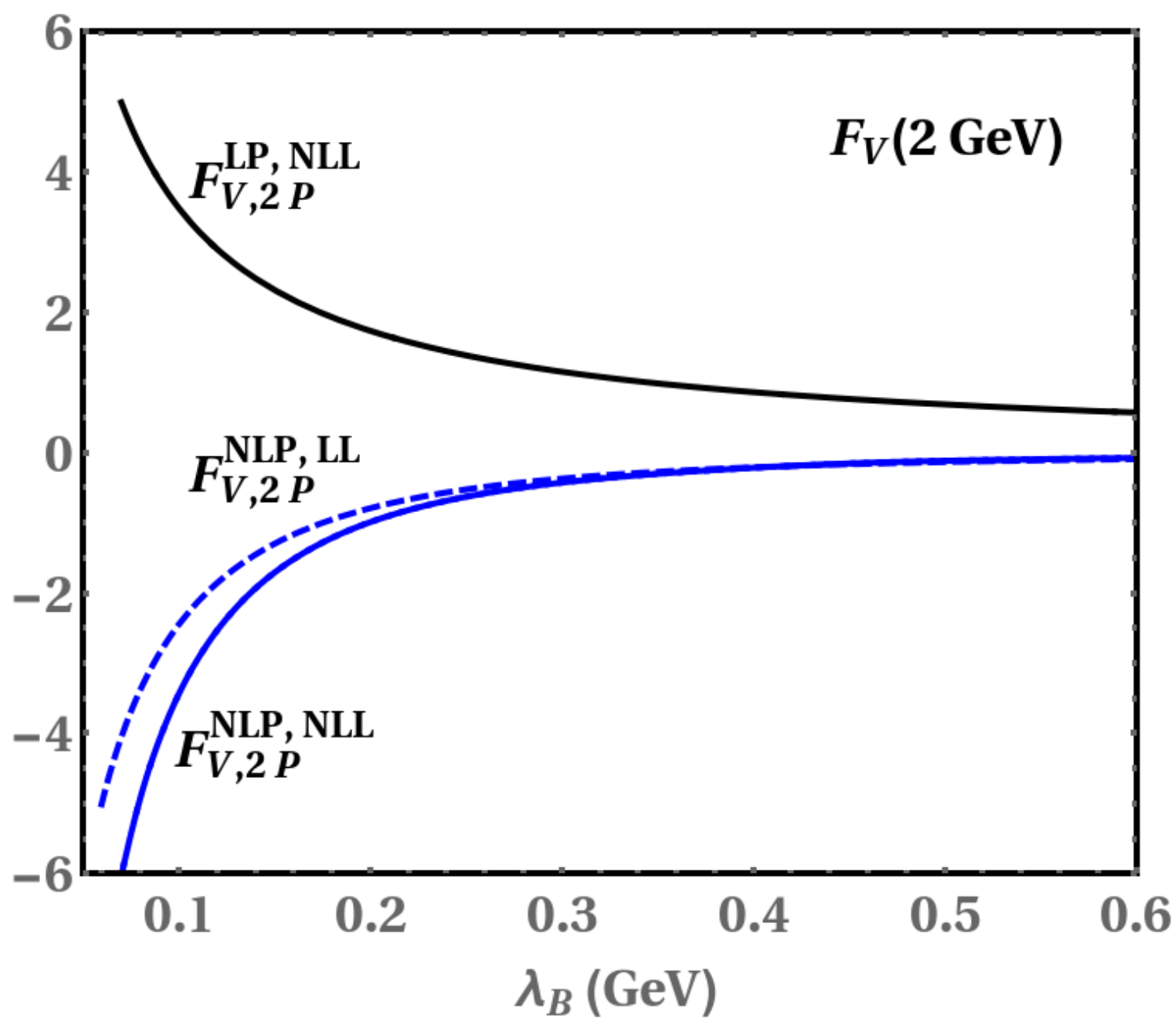}
\end{center}
\caption{Dependence of the leading and subleading power two-particle contributions
to the form factor $F_V(n \cdot p)$ on the inverse moment  $\lambda_B(\mu_0)$
at zero momentum transfer (left panel) and at $n \cdot p= 2 \, {\rm GeV}$ (right panel).
Same conventions as in figure \ref{Breakdown of different contributions for the energy dependence}.}
\label{Breakdown of different contributions for the lambdaB dependence}
\end{figure}

Keeping in mind that we aim at deriving the theory bound for the inverse moment $\lambda_B(\mu_0)$
of the $B$-meson DA $\phi_B^{+}(\omega, \mu_0)$ with the experimental data of the partial
branching fractions of $B \to \gamma \ell \nu$, it is of interest to investigate the $\lambda_B$
dependence of the subleading power corrections to the $B \to \gamma$ form factors.
As can be observed from figure \ref{Breakdown of different contributions for the lambdaB dependence},
the power suppressed two-particle contribution $F_{V, 2P}^{\rm{NLP, NLL}}$ decreases  rapidly
for  $\lambda_B \leq 150 \, {\rm MeV}$ and it leads to a rather sizeable correction to the leading power prediction
of the vector $B \to \gamma$ form factor $F_{V, 2P}^{\rm{LP, NLL}}$ for a  reference value
$\lambda_B(\mu_0)=100 \, {\rm MeV}$:  ${\cal O} (45 \, \%)$ at $n \cdot p=m_B$ and ${\cal O} (100\, \%)$
at $n \cdot p=2 \, {\rm GeV}$. Also, we find that the NLL resummation improved perturbative correction
to the soft two-particle contribution becomes more important numerically with the decrease of $\lambda_B(\mu_0)$:
approximately $(20 \sim 40) \%$ for $ n \cdot p \in  \left [ 2 \, {\rm GeV} \,, m_B \right ]$
with $\lambda_B(\mu_0)=100 \, {\rm MeV}$. We can readily conclude the ``soft" two-particle contribution to the
on-shell $B \to \gamma$ form factors is not effectively suppressed numerically at small $\lambda_B(\mu_0)$
as expected from the power counting analysis in the heavy quark limit.
Moreover, we observe that the subleading power three-particle correction to the $B \to \gamma$
form factors is still insignificant even at $\lambda_B(\mu_0) =100 \, {\rm MeV}$,
approximately ${\cal O} (1 \, \%)$,  compared with the factorizable two-particle contribution.

To understand such ``anomalous" feature of the subleading power two-particle correction,
we first recall that the power counting scheme established above makes use of the canonical behaviour of the
$B$-meson DA $\phi_B^{+}(\omega, \mu_0)$ \cite{Beneke:2000ry}
\begin{eqnarray}
\phi_B^{+}(\omega, \mu_0)  \sim \left\{
\begin{array}{l}
{1 / \Lambda}  \,; \qquad
\omega \sim \Lambda \vspace{0.4 cm} \\
0 \,; \qquad  \hspace{0.5 cm} \omega \gg \Lambda
\end{array}
 \hspace{0.5 cm} \right. ,
\end{eqnarray}
which implies that the inverse moment $\lambda_B(\mu_0)$ scales as $\Lambda$ in consistent with
the generic scaling of the light-quark momentum in the $B$-meson.
However, it would be more appropriate to count the scaling of the inverse moment as
$\lambda_B(\mu_0) \sim \Lambda^2/m_b$ for
$\lambda_B(\mu_0) \leq 100 \, {\rm MeV}$ in the heavy quark limit.
Applying this new power counting scheme leads to
\begin{eqnarray}
F_{V, 2P}^{\rm{LP}} \sim F_{V, 2P}^{\rm{NLP}} \sim
\left ( {m_b \over \Lambda} \right )^{1/2} \,, \qquad
{\rm for} \hspace{0.5 cm} \lambda_B(\mu_0) \sim \Lambda^2/m_b \,,
\end{eqnarray}
which  indicates that the ``soft" two-particle contribution to the
$B \to \gamma$ form factors is of the same power in the heavy quark expansion
as the factorizable effect computed from the QCD factorization approach.
To validate the leading-power factorization formula for the generalized $B \to \gamma^{\ast}$
form factors (\ref{resummation improved factorization formula}), we will therefore only focus on the
inverse moment region $\lambda_B(\mu_0) \geq 200 \, {\rm MeV}$ in accordance with
the power counting $\lambda_B(\mu_0) \sim \Lambda$ in the following analysis,
which implies the desired power counting rule for the ``soft" two-particle correction
\begin{eqnarray}
F_{V, 2P}^{\rm{LP}} \sim \left ( {\Lambda \over m_b} \right )^{1/2} \,, \qquad
F_{V, 2P}^{\rm{NLP}} \sim \left ( {\Lambda \over m_b} \right )^{3/2} \,, \qquad
{\rm for} \hspace{0.5 cm} \lambda_B(\mu_0) \sim \Lambda \,.
\end{eqnarray}

\begin{figure}
\begin{center}
\includegraphics[width=0.45 \textwidth]{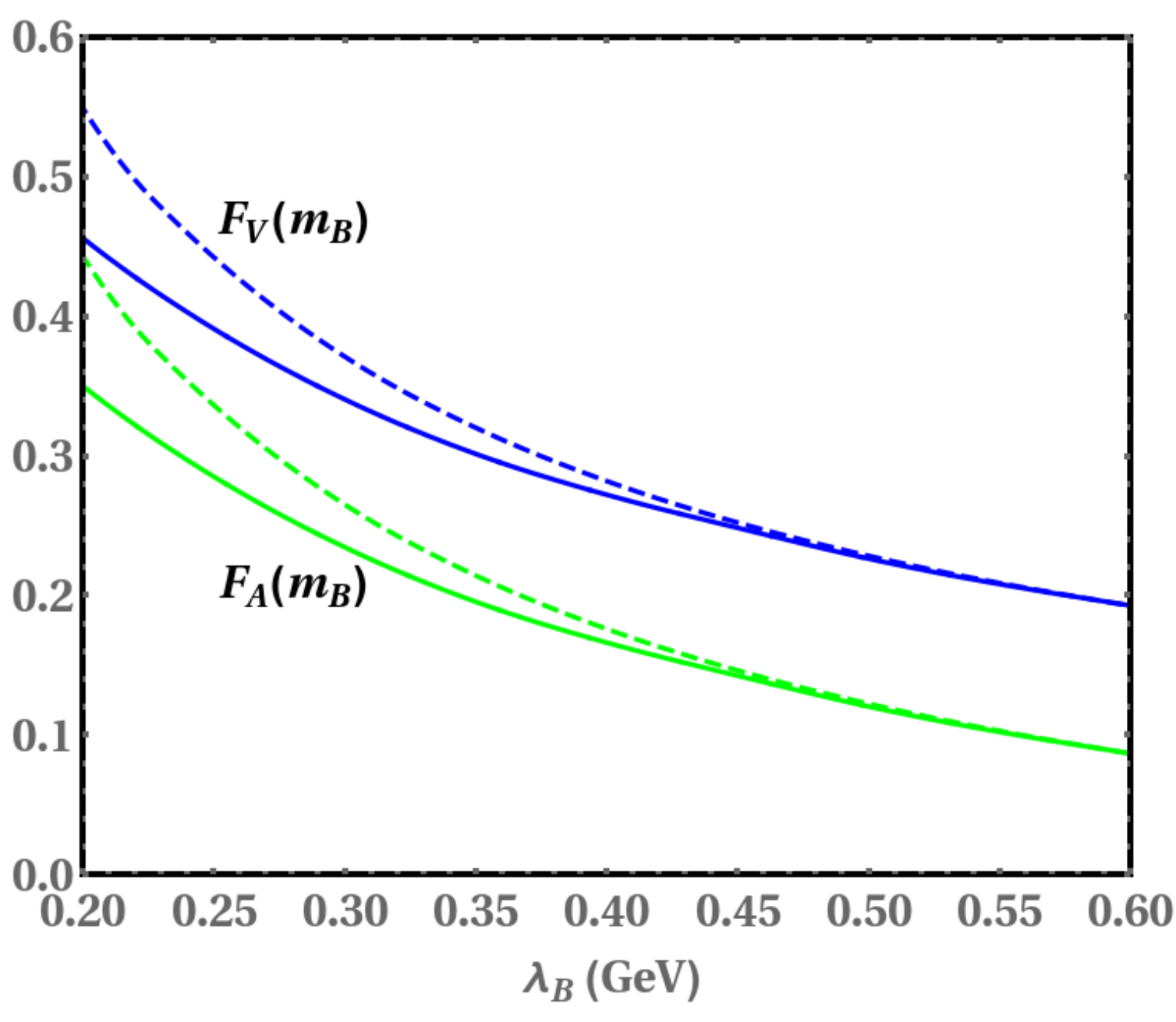}
\includegraphics[width=0.45 \textwidth]{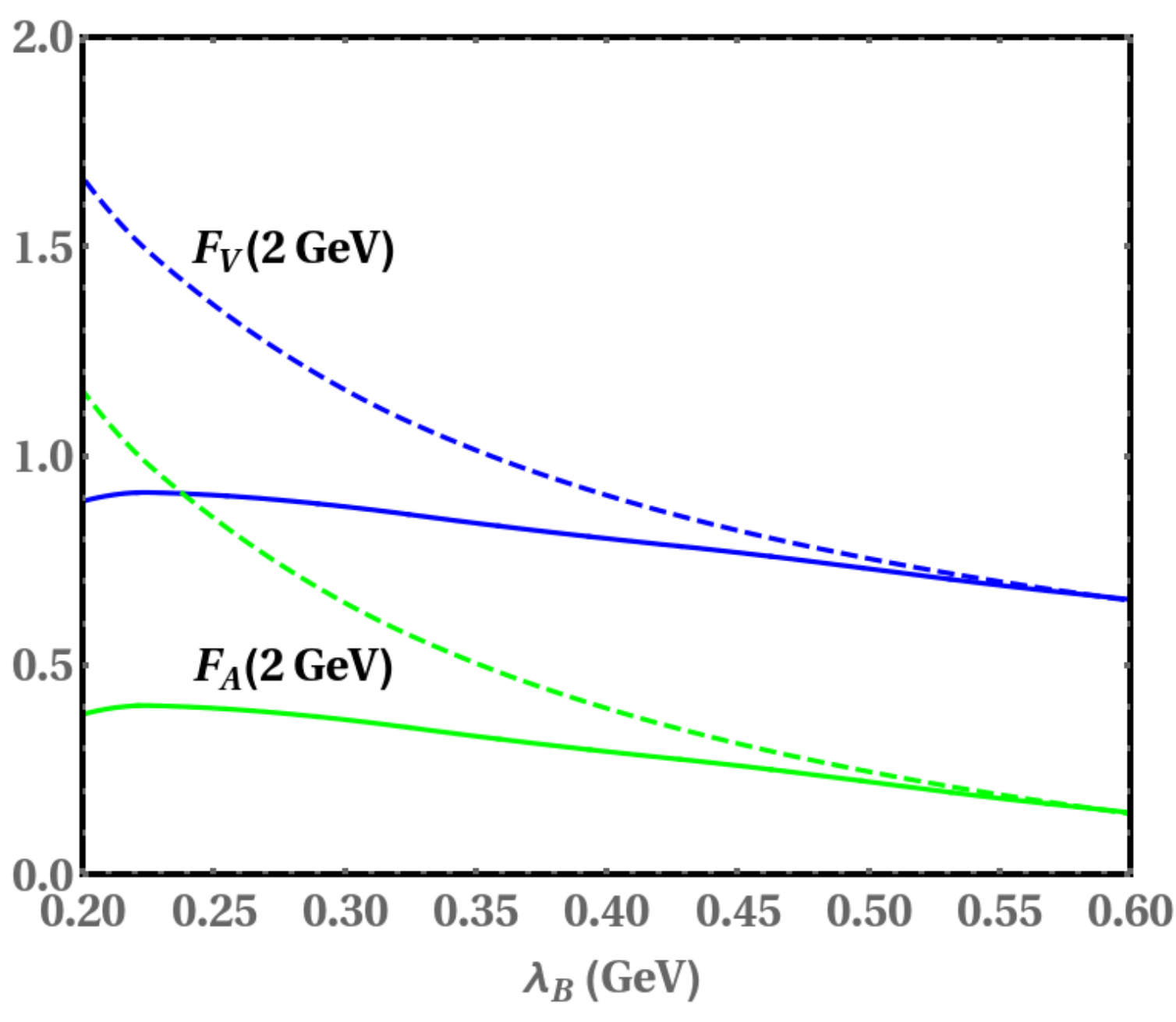}
\end{center}
\caption{Dependence of the form factors $F_V (n \cdot p)$ and $F_A (n \cdot p)$ on the specific model for
the $B$-meson DA $\phi_B^{+}(\omega, \mu)$ at $n \cdot p=m_B$ (left panel) and at $n \cdot p=2 \, {\rm GeV}$ (right panel).
The solid and dashed  blue (green) curves indicate the theory predictions of $F_V $ ($F_A$) from the first and second models
of $\phi_B^{+}$ displayed in (\ref{the first model of the B-meson DA}) and (\ref{the second model of the B-meson DA}), respectively.}
\label{model dependence of B to gamma FFs}
\end{figure}

We turn to investigate phenomenological impacts of  the model dependence of $\phi_B^{+}(\omega, \mu)$
on the theoretical predictions of the $B \to \gamma$ form factors.
It is evident from figure \ref{model dependence of B to gamma FFs} that the form factors
$F_V$ and $F_A$ are insensitive to the specific model of the $B$-meson DA $\phi_B^{+}(\omega, \mu)$
for a  large value of $\lambda_B(\mu_0)$. This can be readily understood from the fact that the leading power
contribution to the $B \to \gamma$ form factors is determined by the inverse-logarithmic moments completely
and the subleading power two-particle and three-particle corrections are both parametrically and numerically
suppressed compared with the leading power effect at large $\lambda_B(\mu_0)$.
The distinct predictions of the $B \to \gamma$ form factors from different nonperturbative
models  of $\phi_B^{+}(\omega, \mu)$  at small $\lambda_B(\mu_0)$, displayed in figure
\ref{model dependence of B to gamma FFs}, imply that soft (end-point) contributions to the form factors
$F_V$ and $F_A$ are both  numerically sizable and heavily dependent on the precise shape of the $\phi_B^{+}(\omega, \mu)$
at small $\omega$, in agreement with a similar observation for the $B \to \pi$ form factors \cite{Wang:2015vgv}.
In particular, the resulting discrepancies for the form factor predictions
due to different parameterizations of  $\phi_B^{+}(\omega, \mu)$ will be further enhanced at
$n \cdot p = 2 \, {\rm GeV}$ due to the raise of power suppressed corrections.

\begin{figure}[t]
\begin{center}
\includegraphics[width=0.60 \textwidth]{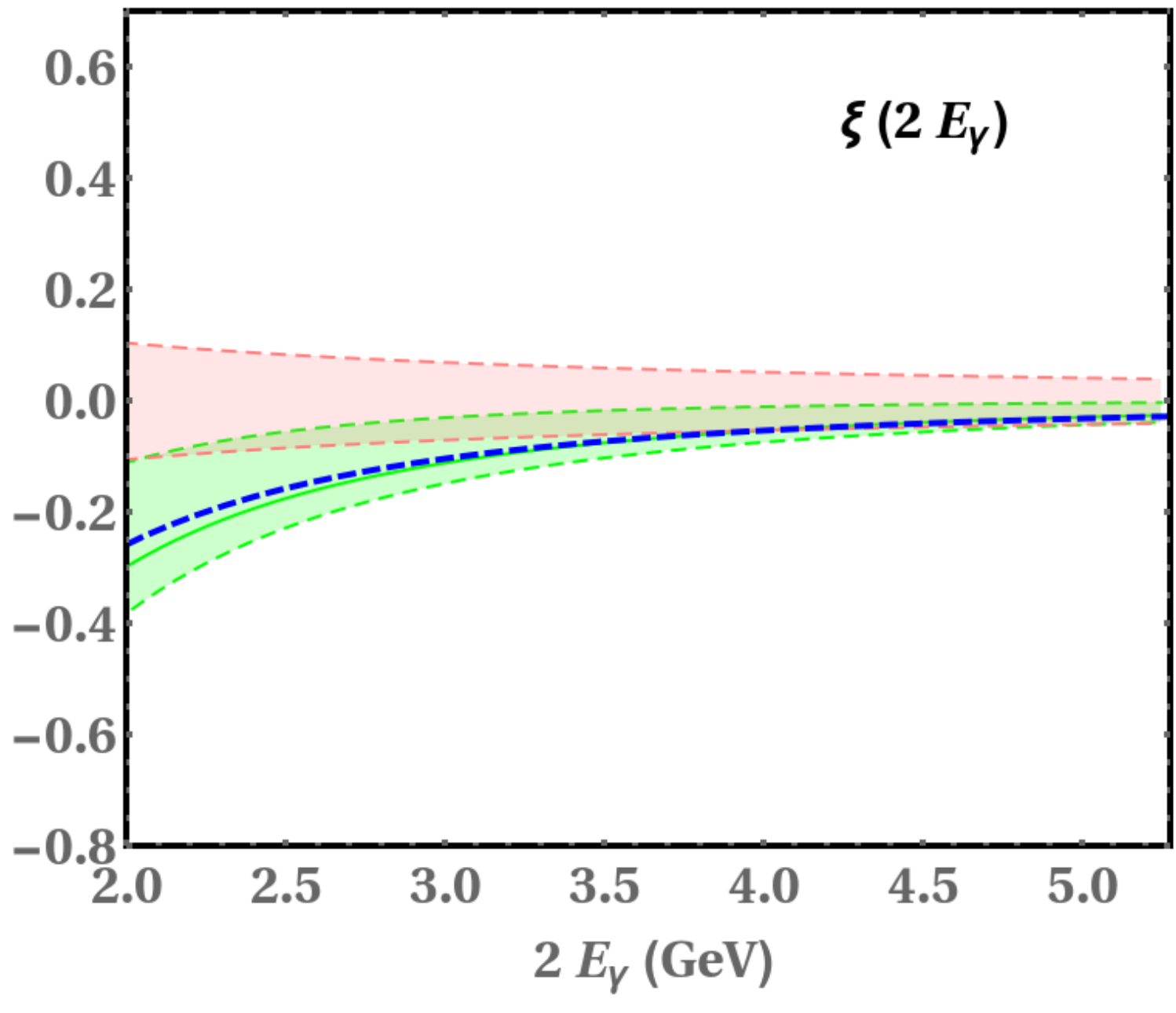}
\end{center}
\caption{The non-local effect due to photon radiation off the up anti-quark
parameterized by the subleading power form factor $\xi(2 \, E_{\gamma})$ (pink band)
\cite{Beneke:2011nf} compared with the sum of the power suppressed two-particle and three-particle
corrections $F_{V, 2P}^{\rm{NLP, NLL}}+F_{V, 3P}^{\rm{LO}}$ computed from the dispersion approach
(green band) with $\lambda_B(\mu_0) =  354 \, {\rm MeV}$  determined in \cite{Wang:2015vgv}.
The blue curve refers to the LL prediction for the soft two-particle contribution
with the default choices of theory inputs.}
\label{comparision of the subleading form factor}
\end{figure}

Now we proceed to perform a  numerical comparison of  the power suppressed
two-particle and three-particle corrections to the $B \to \gamma$
form factors $F_{V, 2P}^{\rm{NLP, NLL}}+F_{V, 3P}^{\rm{LO}}$, computed from the dispersion approach,
and the subleading power  symmetry-conserving form factor $\xi(2 \, E_{\gamma})$ introduced in \cite{Beneke:2011nf}.
In the absence of a detailed analysis of  $\xi(2 \, E_{\gamma})$, a simple model
in compatible with the power counting analysis in the heavy quark limit
\begin{eqnarray}
\xi(2 \, E_{\gamma}) = c \, {f_B \over 2 \, E_{\gamma}} \,
\label{simple model for xi}
\end{eqnarray}
was proposed in \cite{Beneke:2011nf}, assuming the same $E_{\gamma}$ dependence as
the leading power contribution $F_{V, 2P}^{\rm{LP}}(2 \, E_{\gamma})$.
One can readily conclude from figure \ref{comparision of the subleading form factor} that
the nonperturbative parameter $c$ needs to  be significantly larger than one  at
$E_{\gamma} \simeq 1 \, {\rm GeV}$ so that  $\xi(2 \, E_{\gamma})$ can match the non-local
subleading power contributions to  the $B \to \gamma$ form factors numerically, confirming
the observation made in \cite{Braun:2012kp}. In addition, we observe that the photon-energy
dependence of the soft contribution  $F_{V, 2P}^{\rm{NLP}}+F_{V, 3P}$ cannot be well
described by the simple model (\ref{simple model for xi}) particularly for $E_{\gamma} \leq 1.5 \, {\rm GeV}$.

\begin{figure}[t]
\begin{center}
\includegraphics[width=0.60 \textwidth]{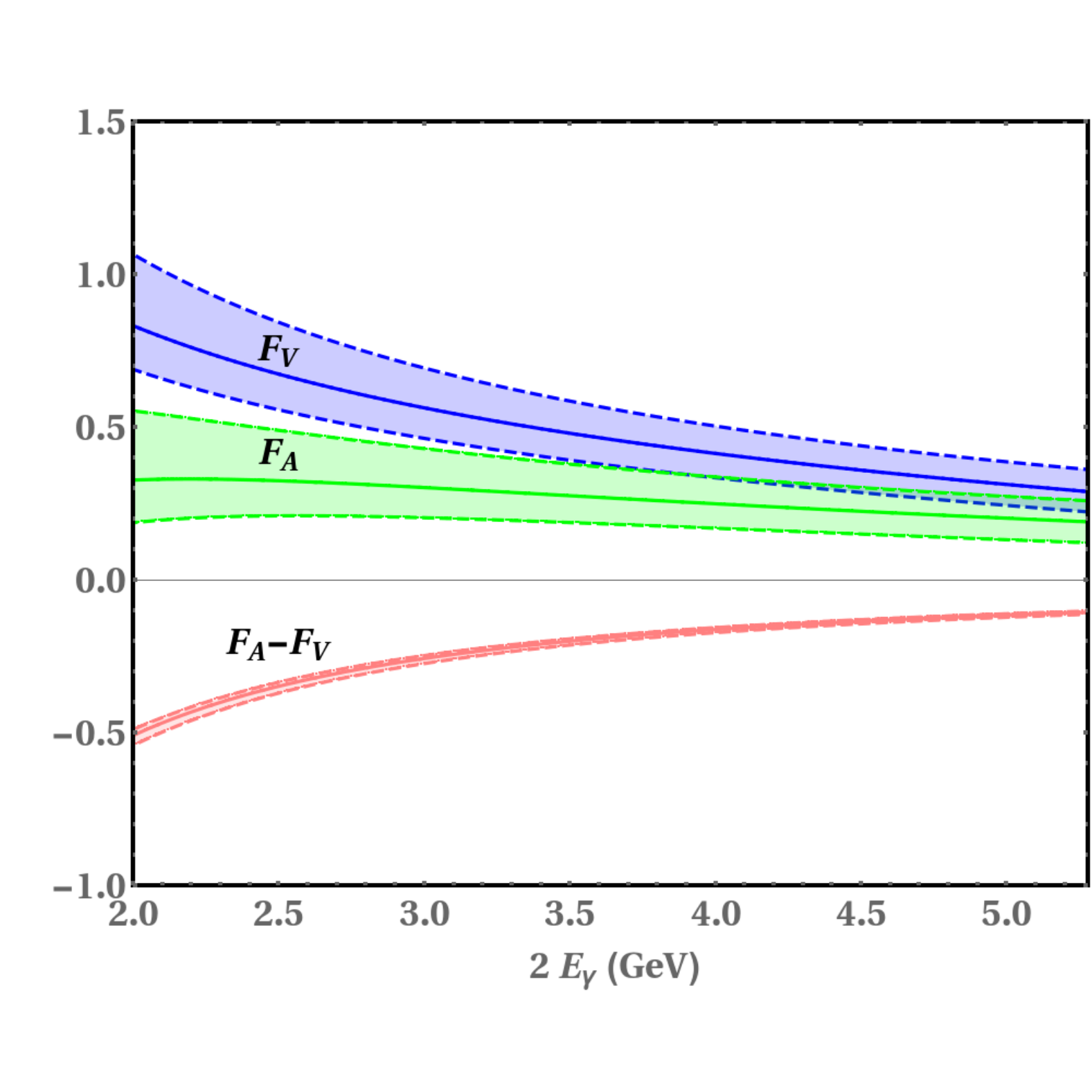}
\end{center}
\caption{The photon-energy dependence of the form factors $F_V(2 \, E_{\gamma})$
and $F_A(2 \, E_{\gamma})$ as well as their difference with $\lambda_B(\mu_0) =  354 \, {\rm MeV}$.
The theory uncertainties from variations of different input parameters are added in quadrature.}
\label{figure: photon-energy dependence of FFs}
\end{figure}

We further present the  main theory predictions for the photon-energy dependence of the $B \to \gamma$
form factors in  figure \ref{figure: photon-energy dependence of FFs}, taking into account the newly computed
power suppressed two-particle and three-particle contributions  $F_{V, 2P}^{\rm{NLP}}+F_{V, 3P}$.
Several comments on the numerical results obtained above are in order.
\begin{itemize}
\item{The dominant theory uncertainties arise from the factorization scale $\mu$, the inverse-logarithmic
moments $\lambda_B(\mu_0)$, $\sigma_1(\mu_0)$ and $\sigma_2(\mu_0)$, as well as the model dependence
of the $B$-meson DA $\phi_B^{+}(\omega, \mu_0)$.   The strong sensitivity of the soft two-particle contribution to the  precise
shape of $\phi_B^{+}(\omega, \mu_0)$ at small $\omega$ is not unexpected by inspecting the analytical expression
of $F_{V, 2P}^{\rm NLP}(n \cdot p)$ in (\ref{NLL 2-particle contribution to form factors}).}
\item{Since the subleading power two-particle and three-particle corrections to the $B \to \gamma$ form factors
preserve the symmetry relation for the leading power contributions due to helicity conservation,
the symmetry-breaking effect still originates from the subleading power local corrections
with the current accuracy \cite{Beneke:2011nf}
\begin{eqnarray}
F_A(n \cdot p)-F_V(n \cdot p)={2 \, f_B \over n \cdot p} \,
\left [Q_{\ell} - {Q_u \, m_B \over n \cdot p} - {Q_b \, m_B \over m_b} \right ]
+ {\cal O}(\alpha_s) \,,
\end{eqnarray}
dependent only on the $B$-meson decay constant $f_B$.
This also explains why the form factor difference only suffers from a very small uncertainty as displayed
in  figure \ref{figure: photon-energy dependence of FFs}, albeit with the large theory uncertainty for the
individual form factor.}
\item {Since the photon-energy dependence of the form factors $F_V$ and $F_A$ is controlled by the nearest poles
in the vector and axial-vector $\left (b \, \bar u \right )$ channels, the vector form factor $F_V$
grows faster than $F_A$ with the increase of $q^2$ (i.e., with the decrease of $E_{\gamma}$) in compatible with
the  prediction presented in figure \ref{figure: photon-energy dependence of FFs}.  }
\end{itemize}

\begin{figure}[t]
\begin{center}
\includegraphics[width=0.45 \textwidth]{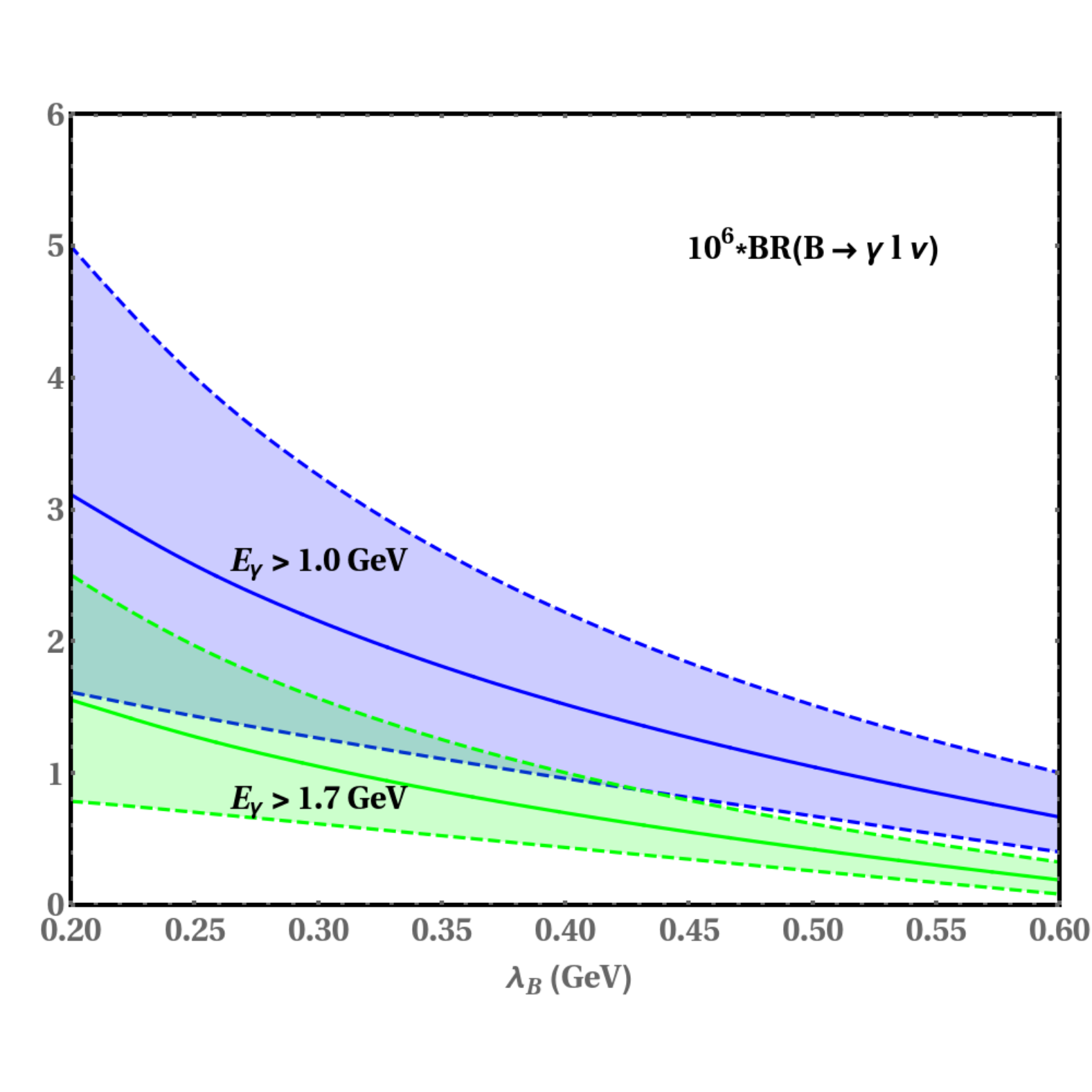} \hspace{0.2 cm}
\includegraphics[width=0.45 \textwidth]{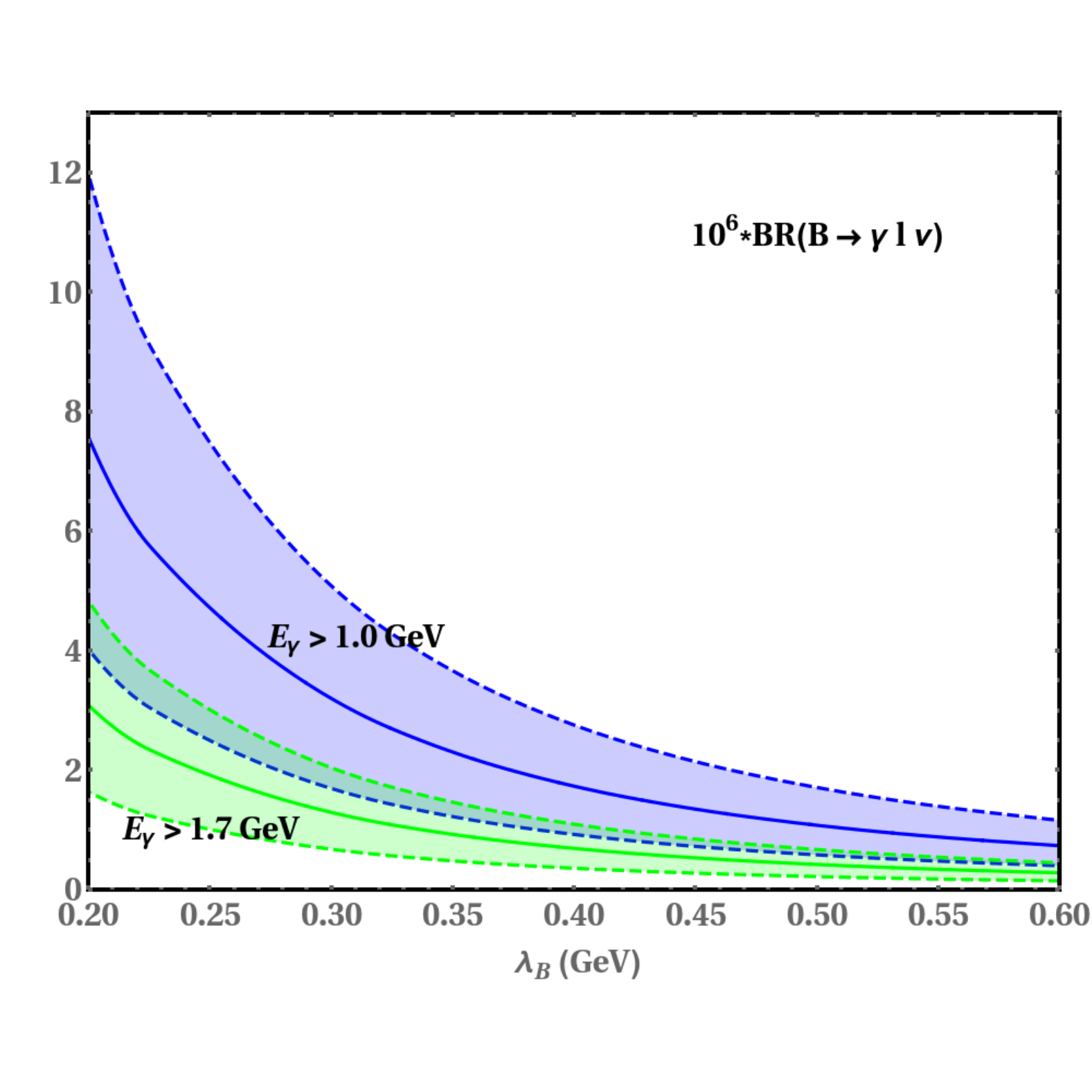}
\end{center}
\caption{The inverse-moment $\lambda_B(\mu_0)$ dependence of the
partial branching  fractions of
${\cal BR} (B \to \gamma \ell \nu, E_{\gamma} \geq E_{\rm cut})$
for $E_{\rm cut}=1 \, {\rm GeV}$ (blue band) and  $E_{\rm cut}=1.7 \, {\rm GeV}$
(green band) with the  model $\phi_{B, \rm I}^{+}(\omega, \mu_0)$
based upon the Grozin-Neubert parametrization (left panel)
and with the  model $\phi_{B, \rm {II}}^{+}(\omega, \mu_0)$
based upon the Braun-Ivanov-Korchemsky parametrization (right panel). }
\label{figure: branching fractions}
\end{figure}

Having at our disposal the theory predictions for the $B \to \gamma$ form factors,
we turn to explore the theory constraint on $\lambda_B(\mu_0)$ from the partial branching fractions
of $B \to \gamma \ell \nu$. Since the factorization formula for the decay amplitude
${\cal A}(B^{-} \to \gamma \, \ell \, \nu )$
was established with the power counting scheme $n \cdot p \equiv 2 \, E_{\gamma} \sim {\cal O}(m_b)$,
the phase-space cut on the photon energy needs to be introduced in the definition of the integrated decay rate
\begin{eqnarray}
\Delta {\cal BR}(E_{\rm cut}) = \tau_{B} \, \int_{E_{\rm cut}}^{m_B/2} \, d \, E_{\rm \gamma} \,\,
\frac{d \, \Gamma}{ d \, E_{\rm \gamma}} \left ( B \to \gamma \ell \nu \right ) \,,
\end{eqnarray}
in order to facilitate the comparison of the experimental measurements from the Belle Collaboration \cite{Heller:2015vvm}
and the theoretical predictions displayed in figure \ref{figure: branching fractions}.
The main observations can be summarized as follows.
\begin{itemize}
\item {Employing the upper limit of the partial branching fraction with  $E_{\rm cut}=1 \, {\rm GeV}$ from the Belle
experiment $\Delta {\cal BR}(1 \, {\rm GeV})< 3.5 \, \times 10^{-6}$, we find that no interesting bound on
$\lambda_B(\mu_0)$ for the Grozin-Neubert model (\ref{the first model of the B-meson DA})
can be deduced from the weak experiment limit, when the subleading power two-particle and three-particle
corrections to the $B \to \gamma$ form factors are taken into account in the theory predictions.
In contrast, applying the formulae for the transition form factors $F_V $ and $F_A$
computed from QCD factorization  \cite{Beneke:2011nf} directly yields a meaningful bound
$\lambda_B(\mu_0) > 217 \, {\rm MeV}$.  The discrepancy can be traced back to the rapidly growing
soft two-particle contribution with the reduction of $\lambda_B(\mu_0)$ as presented in figure
\ref{Breakdown of different contributions for the lambdaB dependence}, which can induce a strong
cancellation between the leading power contributions and the power suppressed effects.
We are therefore led to conclude that the power suppressed two-particle
and three-particle contributions computed in this work are indispensable to the extraction of the inverse moment
$\lambda_B(\mu_0)$ from the radiative leptonic $B \to \gamma \ell \nu $ decay.}
\item {Due to the apparent larger branching fractions of $B \to \gamma \ell \nu$ predicted from the model
$\phi_{B, \rm {II}}^{+}(\omega, \mu_0)$ in (\ref{the second model of the B-meson DA})
at small $\lambda_B(\mu_0)$, the above-mentioned Belle limit yields
a  loose bound $\lambda_B(\mu_0) > 214 \, {\rm MeV}$.
The strong sensitivity of the  extracted bound of $\lambda_B(\mu_0)$ on the
parametrization of the leading-twist $B$-meson DA can be understood from the model-dependence
of $\phi_B^{+}(\omega, \mu)$ on predicting the $B \to \gamma$ form factors displayed
in  figure \ref{model dependence of B to gamma FFs}. Such model dependence in the evaluation of the $B \to \gamma$
form factors will be significantly reduced only for $\lambda_B(\mu_0) \geq 500 \, {\rm MeV}$
where the leading power contribution to the form factors $F_V$ and $F_A$ computed from QCD factorization approach
also turns out to be numerically dominant. Precision measurements of the binned distribution of $B \to \gamma \ell \nu$
from the forthcoming Belle II experiment at KEK are expected to shed light on the information of $\phi_B^{+}(\omega, \mu)$
at small $\omega$. }
\end{itemize}

\section{Conclusion and outlook }
\label{sect:Conc}

Applying the dispersion approach developed in the context of the pion-photon transition form factor,
we computed perturbative QCD corrections to the subleading power soft two-particle contribution of the
$B \to \gamma$ transition form factors, which cannot be addressed directly with the QCD factorization
approach due to the breakdown of light-cone OPE in the end-point region.
To achieve this goal, we first demonstrated QCD factorization for the generalized $B \to \gamma^{\ast}$
form factors with a hard-collinear photon at leading power in $\Lambda/m_b$ using the diagrammatic factorization approach.
Both the hard coefficient and  jet function entering the factorization formulae for the $B \to \gamma^{\ast}$
form factors were determined at one loop explicitly based upon the method of regions.
We further verified that the hard function $C_{\perp}$ is consistent with the perturbative matching coefficient of the
QCD weak current  $\bar u \, \gamma_{\mu \perp} \, (1- \gamma_5)\, b$  in SCET, and the hard-collinear function $J_{\perp}$  reproduces
the jet function involved in the factorization formulae for the on-shell $B \to \gamma$ form factors when setting
$\bar n \cdot p \to 0$. Employing the RG evolution equations in the momentum space, we obtained the NLL resummation
improved factorization formulae for the generalized $B \to \gamma^{\ast}$ form factors at leading power in $\Lambda/m_b$,
which allows one to derive the expression for the soft two-particle correction to the  $B \to \gamma$ form factors
straightforwardly with the standard dispersion relation in the variable $p^2$.
We also mention in passing that the above-mentioned factorization formulae for the $B \to \gamma^{\ast}$ form factors
can be also employed to construct the NLL sum rules for the $B \to \rho$ form factors at large recoil.

Along the same vein, we also constructed the factorization formula for the three-particle contribution
to the generalized $B \to \gamma^{\ast}$ form factors at tree level.
In accordance with the end-point behaviours of the three-particle $B$-meson DAs, we showed that QCD factorization
for the three-particle contribution to the on-shell $B \to \gamma$ form factors is violated due to the rapidity
divergences in the corresponding convolution integrals.
Moreover,  both the ``soft" and ``hard" three-particle corrections to the $B \to \gamma$ form factors
were shown to contribute at the same power in $\Lambda/m_b$ with the aid of the dispersion approach,
in contrast to the two-particle counterparts.
In particular, the newly computed subleading power two-particle and three-particle corrections turn out to
preserve the symmetry relation of the leading power contribution to $F_V$ and $F_A$ as a consequence of the
helicity conservation.

Having at hand the dispersion expressions for the $B \to \gamma$ form factors, we proceeded to explore the phenomenological
impacts of the power suppressed two-particle and three-particle contributions in detail.
Employing the nonperturbative models of the $B$-meson DA $\phi_B^{+}(\omega, \mu_0)$ motivated from the tree-level and the NLO
QCD sum rule computations, we found that perturbative QCD corrections to the soft two-particle contribution can
give rise to  $(10 \sim 20) \%$ shift to the tree-level prediction at $\lambda_B(\mu_0)=354 \, {\rm MeV}$,
and the three-particle correction to the $B \to \gamma$ form factors at leading order in $\alpha_s$ was found
to be of ${\cal O} (1 \%)$ numerically with the exponential model of the three-particle DAs and with the same value of
the inverse moment. However,  the soft two-particle correction to the $B \to \gamma$ form factors can be significantly
enhanced for $\lambda_B(\mu_0) \leq 150 \, {\rm MeV}$ and it yields a strong cancellation against
the leading power contributions computed in QCD factorization.
We further argued that the ``anomalous" soft two-particle contribution at small $\lambda_B(\mu_0)$ can be understood
from the power counting analysis of the analytical expression  (\ref{NLL 2-particle contribution to form factors})
with an appropriate scaling $\lambda_B \sim \Lambda^2/m_b$ in this regard.
Numerically the subleading power two-particle and three-particle contributions to the $B \to \gamma$ form factors
were evaluated to be considerably greater than the power suppressed symmetry-conserving form factor $\xi(2 \, E_{\gamma})$,
estimated from the simple phenomenological model (\ref{simple model for xi}), at $E_{\gamma} \simeq 1\, {\rm GeV}$.
Our main theory predictions for the form factors $F_V$ and $F_A$ including the subleading power contributions from
the two-particle DA $\phi_B^{+}$ at NLL and from the three-particle DAs at tree level were presented in
figure \ref{figure: photon-energy dependence of FFs}.
With the theory predictions for the $B \to \gamma$ form factors at hand, we proceeded with computing
the integrated branching fractions of $B \to \gamma \ell \nu$ with the phase-space cut on the photon energy
$E_{\gamma} \geq E_{\rm cut}$. The theory constraint  of the inverse moment $\lambda_B$ derived from the
recent Belle data on ${\cal BR}(B \to \gamma \ell \nu)$ was found to be sensitive to the specific model
of $\phi_B^{+}$ adopted in the evaluation of the form factors $F_V$ and $F_A$, since the subleading power
soft two-particle  correction is not sufficiently suppressed numerically at small $\lambda_B$
and dependent on the precise shape of $\phi_B^{+}$ at small $\omega$.
Remarkably, no interesting bound on the inverse moment $\lambda_B$
can be derived,  with the model  $\phi_{B, \rm I}^{+}(\omega, \mu_0)$ in (\ref{the first model of the B-meson DA}),
from the inconclusive Belle measurement  $\Delta {\cal BR}(1 \, {\rm GeV})< 3.5 \, \times 10^{-6}$,
when the subleading power two-particle and three-particle corrections are taken into account.
In contrast, employing an alternative model based on the Braun-Ivanov-Korchemsky  parametrization
(\ref{the second model of the B-meson DA}) would yield a
weak bound $\lambda_B(\mu_0) > 214 \, {\rm MeV}$ from the Belle data, due to the  substantially enhanced predictions for
the branching fractions of $B \to \gamma \ell \nu$ at small $\lambda_B(\mu_0)$.

Exploring the strong interaction dynamics of the radiative leptonic $B \to \gamma \ell \nu$ decay beyond this work
can be pursued in different directions. First,  it would be of  interest to investigate the factorization property
of the subleading power form factor $\xi(2 \, E_{\gamma})$ in QCD, and then to build up the relation between the non-local
subleading power corrections computed from the dispersion approach and $\xi(2 \, E_{\gamma})$ expressed in terms of the
higher-twist $B$-meson DAs. Second, calculating the yet higher-twist corrections to the $B \to \gamma$ form factors
from the four-particle $B$-meson DAs in the framework of  the dispersion approach
will be helpful to clarify  whether they are indeed suppressed by  one power
of $\Lambda/m_b$ due to the mismatch between the twist expansion and the power expansion, and to verify whether
the non-local higher-twist contributions generate the symmetry-breaking effect between $F_V$ and $F_A$ as
observed from the sum rule approach with the photon DAs.
Third, extending the current analysis by computing  perturbative corrections to
the three-particle contributions of the generalized $B \to \gamma^{\ast}$ form factors
will deepen our understanding towards QCD factorization for the subleading power contributions
in exclusive $B$-meson decays, and more important, such computations will be essential to construct
the NLL sum rules for $B \to \rho$ form factors even at leading power in $\Lambda/m_b$.
To summarize, we believe that precision QCD calculations of the radiative $B \to \gamma \ell \nu$ decay
are sufficiently interesting on both the conceptual and  phenomenological aspects.

\subsection*{Acknowledgements}

The author is grateful to Martin Beneke and Vladimir Braun for illuminating discussions,
and to Vladimir Braun for valuable comments on the manuscript.


\appendix


\section{Spectral representations}
\label{app:spectral resp}

Here we collect the dispersion representations of various convolution integrals
involved in the factorization formulae for the generalized $B \to \gamma^{\ast}$
form factors presented in (\ref{resummation improved factorization formula}).
In particular, we confirm the following spectral representations by verifying  the
corresponding dispersion integrals  manifestly.

\begin{eqnarray}
&& {1 \over \pi} \, {\rm Im}_{\omega^{\prime}} \, \int_0^{\infty} \,  \,
\, \frac{d \omega}{\omega-\omega^{\prime}-i0} \,
\ln^2{\mu^2 \over n \cdot p \, (\omega-\omega^{\prime})} \,\, \phi_B^{+}(\omega, \mu) \nonumber \\
&& = \int_0^{\infty} \, d \omega \,
\left [ {2 \, \theta(\omega^{\prime}-\omega) \over \omega - \omega^{\prime}}  \,
\ln {\mu^2 \over n \cdot p \, (\omega^{\prime} - \omega)}   \right ]_{\oplus}
\,\, \phi_B^{+}(\omega, \mu)
+ \left [ \ln {\mu^2 \over n \cdot p \,  \omega^{\prime} }
- {\pi^2 \over 3} \right ] \phi_B^{+}(\omega^{\prime}, \mu) \,, \\
&& {1 \over \pi} \, {\rm Im}_{\omega^{\prime}} \, \int_0^{\infty} \,  \,
\, \frac{d \omega}{\omega-\omega^{\prime}-i0} \,\, { \omega^{\prime} \over \omega } \,\,
\ln {\omega^{\prime} - \omega \over \omega^{\prime}} \,\,
\ln {\mu^2 \over - n \cdot p \, \omega^{\prime}} \,\, \phi_B^{+}(\omega, \mu) \nonumber \\
&& = -{\omega^{\prime} \over 2} \,\,  \bigg \{ \int_0^{\infty} d \omega \,
\ln^2 \bigg|{\omega - \omega^{\prime}  \over \omega^{\prime} } \bigg|  \,\,
{d \over d \omega} \, {\phi_B^{+}(\omega^{\prime}, \mu) \over \omega}  \nonumber \\
&& \hspace{1.5 cm} + \,  \int_{\omega^{\prime}}^{\infty} d \omega \,  \,
\left [ 2 \, \ln{\mu^2 \over n \cdot p \, \omega^{\prime}} \,
\ln {\omega - \omega^{\prime}  \over \omega^{\prime} }  - \pi^2 \right ] \,
{d \over d \omega} \, {\phi_B^{+}(\omega^{\prime}, \mu) \over \omega}  \bigg \} \,, \\
&& {1 \over \pi} \, {\rm Im}_{\omega^{\prime}} \, \int_0^{\infty} \,  \,
\, \frac{d \omega}{\omega-\omega^{\prime}-i0} \,\, { \omega^{\prime} \over \omega } \,\,
\ln {\omega^{\prime} - \omega \over \omega^{\prime}} \,\,
\ln {\mu^2 \over  n \cdot p \, (\omega - \omega^{\prime})} \,\, \phi_B^{+}(\omega, \mu) \nonumber \\
&& = \omega^{\prime} \, \bigg \{ \int_0^{\infty} \, d \omega \,
\left [ { \theta(\omega^{\prime}  - \omega) \over \omega - \omega^{\prime}} \,
\ln { \omega^{\prime} - \omega \over \omega^{\prime}} \right ]_{\oplus} \,
\, {\phi_B^{+}(\omega^{\prime}, \mu) \over \omega}  \nonumber \\
&& \hspace{1.0 cm} + \, {1 \over 2} \, \int_{\omega^{\prime}}^{\infty} \, d \omega \,
\left [ \ln^2 {\mu^2 \over  n \cdot p \, (\omega - \omega^{\prime})}
- \ln^2 {\mu^2 \over  n \cdot p \,  \omega^{\prime}} + {\pi^2 \over 3}  \right ] \,
{d \over d \omega} \, {\phi_B^{+}(\omega^{\prime}, \mu) \over \omega}  \bigg \}  \,, \\
&& {1 \over \pi} \, {\rm Im}_{\omega^{\prime}} \, \int_0^{\infty} \,  \,
\, \frac{d \omega}{\omega-\omega^{\prime}-i0} \,\, { \omega^{\prime} \over \omega } \,\,
\ln {\omega^{\prime} - \omega \over \omega^{\prime}} \,\, \phi_B^{+}(\omega, \mu) \nonumber \\
&& = - \omega^{\prime} \,  \int_{\omega^{\prime} }^{\infty} \, d \omega \,
\ln {\omega - \omega^{\prime} \over \omega^{\prime} } \,\,
{d \over d \omega } \, {\phi_B^{+}(\omega, \mu) \over \omega} \,.
\end{eqnarray}



\begin{thebibliography}{99}



\bibitem{Korchemsky:1999qb}
  G.~P.~Korchemsky, D.~Pirjol and T.~M.~Yan,
  Phys.\ Rev.\ D {\bf 61} (2000) 114510
  [hep-ph/9911427].





\bibitem{DescotesGenon:2002mw}
  S.~Descotes-Genon and C.~T.~Sachrajda,
  Nucl.\ Phys.\ B {\bf 650} (2003) 356
  [hep-ph/0209216].



\bibitem{Lunghi:2002ju}
  E.~Lunghi, D.~Pirjol and D.~Wyler,
  Nucl.\ Phys.\ B {\bf 649} (2003) 349
  [hep-ph/0210091].








\bibitem{Bosch:2003fc}
  S.~W.~Bosch, R.~J.~Hill, B.~O.~Lange and M.~Neubert,
  Phys.\ Rev.\ D {\bf 67} (2003) 094014
  [hep-ph/0301123].






\bibitem{Beneke:2011nf}
  M.~Beneke and J.~Rohrwild,
  Eur.\ Phys.\ J.\ C {\bf 71} (2011) 1818
  [arXiv:1110.3228 [hep-ph]].





\bibitem{Braun:2012kp}
  V.~M.~Braun and A.~Khodjamirian,
  Phys.\ Lett.\ B {\bf 718} (2013) 1014
  [arXiv:1210.4453 [hep-ph]].




\bibitem{Bauer:2000yr}
  C.~W.~Bauer, S.~Fleming, D.~Pirjol and I.~W.~Stewart,
  Phys.\ Rev.\ D {\bf 63} (2001) 114020
  [hep-ph/0011336].





\bibitem{Bauer:2001yt}
  C.~W.~Bauer, D.~Pirjol and I.~W.~Stewart,
  Phys.\ Rev.\ D {\bf 65} (2002) 054022
  [hep-ph/0109045].





\bibitem{Beneke:2002ph}
  M.~Beneke, A.~P.~Chapovsky, M.~Diehl and T.~Feldmann,
  Nucl.\ Phys.\ B {\bf 643} (2002) 431
  [hep-ph/0206152].





\bibitem{Beneke:1997zp}
  M.~Beneke and V.~A.~Smirnov,
  Nucl.\ Phys.\ B {\bf 522} (1998) 321
  [hep-ph/9711391].






\bibitem{Wang:2015vgv}
  Y.~M.~Wang and Y.~L.~Shen,
  Nucl.\ Phys.\ B {\bf 898} (2015) 563
  [arXiv:1506.00667 [hep-ph]].





\bibitem{Wang:2015ndk}
  Y.~M.~Wang and Y.~L.~Shen,
  JHEP {\bf 1602} (2016) 179
  [arXiv:1511.09036 [hep-ph]].





\bibitem{Khodjamirian:1995uc}
  A.~Khodjamirian, G.~Stoll and D.~Wyler,
  Phys.\ Lett.\ B {\bf 358} (1995) 129
  [hep-ph/9506242].






\bibitem{Eilam:1995zv}
  G.~Eilam, I.~E.~Halperin and R.~R.~Mendel,
  Phys.\ Lett.\ B {\bf 361} (1995) 137
  [hep-ph/9506264].




\bibitem{Ball:2003fq}
  P.~Ball and E.~Kou,
  JHEP {\bf 0304} (2003) 029
  [hep-ph/0301135].





\bibitem{Agaev:2010aq}
  S.~S.~Agaev, V.~M.~Braun, N.~Offen and F.~A.~Porkert,
  Phys.\ Rev.\ D {\bf 83} (2011) 054020
  [arXiv:1012.4671 [hep-ph]].






\bibitem{Grinstein:2000pc}
  B.~Grinstein and D.~Pirjol,
  Phys.\ Rev.\ D {\bf 62} (2000) 093002
  [hep-ph/0002216].






\bibitem{Khodjamirian:2001ga}
  A.~Khodjamirian and D.~Wyler,
  In ``Gurzadyan, V.G. (ed.) et al.: From integrable models to gauge theories" 227-241
  [hep-ph/0111249].







\bibitem{Khodjamirian:1997tk}
  A.~Khodjamirian,
  Eur.\ Phys.\ J.\ C {\bf 6} (1999) 477
  [hep-ph/9712451].





\bibitem{Grozin:1996pq}
  A.~G.~Grozin and M.~Neubert,
  Phys.\ Rev.\ D {\bf 55} (1997) 272
  [hep-ph/9607366].



\bibitem{Beneke:2000wa}
  M.~Beneke and T.~Feldmann,
  Nucl.\ Phys.\ B {\bf 592} (2001) 3
  [hep-ph/0008255].





\bibitem{Beneke:2005gs}
  M.~Beneke and D.~S.~Yang,
  Nucl.\ Phys.\ B {\bf 736} (2006) 34
  [hep-ph/0508250].






\bibitem{Bauer:2000ew}
  C.~W.~Bauer, S.~Fleming and M.~E.~Luke,
  Phys.\ Rev.\ D {\bf 63} (2000) 014006
  [hep-ph/0005275].





\bibitem{Beneke:2004rc}
  M.~Beneke, Y.~Kiyo and D.~S.~Yang,
  Nucl.\ Phys.\ B {\bf 692} (2004) 232
  [hep-ph/0402241].





\bibitem{Lange:2003ff}
  B.~O.~Lange and M.~Neubert,
  Phys.\ Rev.\ Lett.\  {\bf 91} (2003) 102001
  [hep-ph/0303082].





\bibitem{Braun:2003wx}
  V.~M.~Braun, D.~Y.~Ivanov and G.~P.~Korchemsky,
  Phys.\ Rev.\ D {\bf 69} (2004) 034014
  [hep-ph/0309330].




\bibitem{Balitsky:1987bk}
  I.~I.~Balitsky and V.~M.~Braun,
  Nucl.\ Phys.\ B {\bf 311} (1989) 541.




\bibitem{Kawamura:2001jm}
  H.~Kawamura, J.~Kodaira, C.~F.~Qiao and K.~Tanaka,
  Phys.\ Lett.\ B {\bf 523} (2001) 111
   Erratum: [Phys.\ Lett.\ B {\bf 536} (2002) 344]
  [hep-ph/0109181].




\bibitem{Geyer:2005fb}
  B.~Geyer and O.~Witzel,
  Phys.\ Rev.\ D {\bf 72} (2005) 034023
  [hep-ph/0502239].






\bibitem{Braun:2015pha}
  V.~M.~Braun, A.~N.~Manashov and N.~Offen,
  Phys.\ Rev.\ D {\bf 92} (2015)   074044
  [arXiv:1507.03445 [hep-ph]].





\bibitem{Khodjamirian:2006st}
  A.~Khodjamirian, T.~Mannel and N.~Offen,
  Phys.\ Rev.\ D {\bf 75} (2007) 054013
  [hep-ph/0611193].






\bibitem{Heller:2015vvm}
  A.~Heller {\it et al.} [Belle Collaboration],
  Phys.\ Rev.\ D {\bf 91} (2015)  112009
  [arXiv:1504.05831 [hep-ex]].







\bibitem{Feldmann:2014ika}
  T.~Feldmann, B.~O.~Lange and Y.~M.~Wang,
  Phys.\ Rev.\ D {\bf 89} (2014)   114001
  [arXiv:1404.1343 [hep-ph]].






\bibitem{Nishikawa:2011qk}
  T.~Nishikawa and K.~Tanaka,
  Nucl.\ Phys.\ B {\bf 879} (2014) 110
  [arXiv:1109.6786 [hep-ph]].






\bibitem{Duplancic:2008ix}
  G.~Duplancic, A.~Khodjamirian, T.~Mannel, B.~Melic and N.~Offen,
  JHEP {\bf 0804} (2008) 014
  [arXiv:0801.1796 [hep-ph]].






\bibitem{Beneke:2014pta}
  M.~Beneke, A.~Maier, J.~Piclum and T.~Rauh,
  Nucl.\ Phys.\ B {\bf 891} (2015) 42
  [arXiv:1411.3132 [hep-ph]].





\bibitem{Beneke:2000ry}
  M.~Beneke, G.~Buchalla, M.~Neubert and C.~T.~Sachrajda,
  Nucl.\ Phys.\ B {\bf 591} (2000) 313
  [hep-ph/0006124].













\end{thebibliography}
\end{document}